\documentclass[11pt]{article}
\usepackage{euscript}

\usepackage{latexsym}
\usepackage{amsmath, amssymb,graphics}
\usepackage[all,2cell]{xy}
\UseTwocells

\usepackage{wrapfig}
\usepackage{graphicx}

\usepackage{color}
\usepackage{tikz}
\usepackage{animate}
\usepackage{calc}
\usepackage{hyperref}
\hypersetup{pdfpagemode=FullScreen}
\newcounter{m} 
\usepackage{fullpage}

\newtheorem{theorem}{Theorem}
\newtheorem{definition}[theorem]{Definition}

\renewcommand{\(}{\begin{equation*}}
\renewcommand{\)}{\end{equation*}}
\newcommand{\bea}{\begin{eqnarray*}}
\newcommand{\eea}{\end{eqnarray*}}
\newcommand{\R}{{\mathbb R}}
\newcommand{\C}{{\mathbb C}}
\newcommand{\Z}{{\mathbb Z}}
\newcommand{\Q}{{\mathbb Q}}
\newcommand{\D}{{\mathbb D}}

\newcommand{\cF}{\ensuremath{\mathcal F}}

\newcommand{\cL}{\ensuremath{\mathcal L}}

\newcommand{\bo}{\raise-1mm\hbox{\Large$\Box$}}              
\def\H{\ensuremath{\ES{H}}}
\def\i{\ensuremath{\dot\imath}}
\def\O{\ensuremath{{\cal O}}}

\newcommand{\cS}{{\mathcal S}}

\def\H{\ensuremath{\ES{H}}}
\def\i{\ensuremath{\dot\imath}}
\def\O{\ensuremath{{\cal O}}}

\newcommand{\beq}{\begin{equation}}
\newcommand{\eeq}{\end{equation}}

\newcommand{\bb}[1]{\ensuremath{\mathbb{#1}}}

\numberwithin{equation}{section}

\renewcommand{\(}{\begin{equation}}
\renewcommand{\)}{\end{equation}}

\def\F{{\mathbb F}}
\def\H{{\mathbb H}}

\def\R{{\mathbb R}}
\def\Z{{\mathbb Z}}
\def\Q{{\mathbb Q}}
\def\C{{\mathbb C}}

\def\1{{\bf 1}}

\def\<{\langle}
\def\>{\rangle}

\def\O{{\cal O}}

\numberwithin{equation}{section}

\renewcommand{\(}{\begin{equation}}
\renewcommand{\)}{\end{equation}}

\begin{document}

\begin{titlepage}


\vspace{2em}
\def\thefootnote{\fnsymbol{footnote}}

\begin{center}
{\Large\bf 
Framed M-branes, corners, and topological invariants
}
\end{center}
\vspace{1em}

\begin{center}
\large Hisham Sati 
\footnote{e-mail: {\tt
hsati@nyu.edu}}
\end{center}

\begin{center}
Division of Science and Mathematics\\
New York University\\
Saadiyat, Abu Dhabi, UAE 
\end{center}

\vspace{0em}
\begin{abstract}

We uncover and highlight relations between the M-branes in M-theory and various
topological invariants: the Hopf invariant over $\Q$, $\Z$ and $\Z_2$, the Kervaire invariant, 
the $f$-invariant, and the $\nu$-invariant. This requires either a framing or 
 a corner structure. 
The canonical framing provides a minimum for the classical action
and the change of framing encodes the structure of the action and possible anomalies. 
We characterize the flux quantization condition on the C-field and 
the topological action of the M5-brane
 via the Hopf invariant, and 
the dual of the C-field as (a refinement of) an element of Hopf invariant two.
In the signature formulation, 
the contribution to the M-brane effective action 
is given by the Maslov index of the corner. 
The Kervaire invariant implies that the effective action of the M5-brane 
is quadratic. Our study leads to viewing 
 the self-dual string, which is  the boundary of the M2-brane on the M5-brane 
worldvolume, as 
 a string theory in the sense of cobordism of manifolds with corners. 
We show that the dynamics of the C-field and its dual are encoded 
in  unified way in the 4-sphere, which 
suggests  the corresponding spectrum as the generalized cohomology theory
describing the fields. The effective action of the corner is captured by the $f$-invariant, which 
is an invariant at chromatic level two. 
Finally, considering M-theory on manifolds with $G_2$ holonomy 
we show that the  canonical $G_2$ structure minimizes the topological 
part of the M5-brane action. This is done via the $\nu$-invariant 
and a variant that we introduce related to the one-loop polynomial.

\end{abstract}

\tableofcontents

\end{titlepage}

\section{Introduction and statement of results}
\label{Sec intro}

The branes in M-theory, i.e. the M-branes, possess rich geometric and topological structures. 
Examples of such structures have been uncovered recently
\cite{tcu}\cite{tcu2}\cite{top1}\cite{top2}. 
In this paper we propose and highlight explicit relations to various topological invariants,
including: the mod 2 and the integral Hopf invariants, the Kervaire invariant, 
the $f$-invariant \cite{Lau}, and the (newly constructed) $\nu$-invariant \cite{CN}. 
A relation to the Hopf invariant at the rational level was also proposed
in \cite{Int}, and the connection of the M5-brane to the 
Kervaire invariant is implicit in the work of Hopkins-Singer
\cite{HS}, which is a precursor to the solution of the Kervaire invariant 
problem \cite{HHR}; a description of the effective action of the M5-brane is 
given in \cite{HS} (along the lines of \cite{Eff}).
The $f$-invariant appears 
in \cite{F} in the description of anomalies in heterotic string theory. 
Combining that with the approach of \cite{boundary}\cite{corner} leads naturally to our discussion
on the relation of the M-branes to the $f$-invariant. This generalizes previous
connections to the Adam's $e$-invariant and the Atiyah-Patodi-Singer 
$\eta$-invariant
\cite{String} \cite{tcu}. 
Recently, a description  of M-branes via higher geometry and higher Chern-Simons theory was
given in \cite{FSS1} \cite{FSS2}  \cite{cup} \cite{up}.

\vspace{3mm}
Making the connections to the above  topological invariants requires certain
 topological structures to be admitted by the M-branes. 
For example, the connection to the Kervaire invariant requires the M-brane
to have a framing, that is, a trivialization of the tangent bundle (see \cite{tcu}). 
Closely related is the notion of String structure and its variations, such as Atiyah's 
2-framing, captured essentially by the C-field $C_3$ and its field strength $F_4$
 \cite{String} \cite{tcu}.
On the other hand, the connection to the $f$-invariant requires a corner
structure on the M-branes. 
For the M5-brane the corner will be the 
six-dimensional worldvolume itself, while for 
 the M2-brane the corner will be either the three-dimensional 
 worldvolume or the boundary 
given by the self-dual string. Furthermore, requiring a $G_2$ holonomy structure
on the 7-dimensional extension of the M5-brane worldvolume leads to 
the direct connection to the $\nu$-invariant. 
We will illustrate these points in detail. 
The discussion will naturally lead along the way to a twisted notion of 
$p_1$-structure (see Def. \ref{def}).

\vspace{3mm}
Once the worldvolumes are endowed with geometric and topological structures, 
it is natural to consider the space of such structures and to investigate whether 
the physical entities depend on the specific structure chosen. It then makes sense
to ask whether there is a distinguished such structure in that set. 
For instance, the space of Spin structures and the space of framings are both affine spaces, and 
hence have no natural base points. However, one can distinguish specific such 
structures which minimize some corresponding invariant; these are the {\it canonical} structures. 
An example of the minimization of the volume of the M2-brane, captured by the String structure, is 
described in \cite{tcu}. Analogously, we consider a distinguished $\nu$-invariant,
which characterizes a minimum for (a properly interpreted)  topological part of the action of the M5-brane. 
The M2-brane will be described either in three dimensions or by moving  up/down by one dimension.
The M5-brane will be described either in six dimensions or by moving up by one or two dimensions. 
Therefore, the relevant dimensions are 2, 3, and 4 for the M2-brane and 6, 7, and 8
for the M5-brane.

\vspace{3mm}
The main content of this paper can be summarized in the following two theorems,
on the relation of the M2-brane and the M5-brane, respectively, to various
topological invariants. Some of the statements might be known to the 
experts but we have not seen them in print, and we believe that, at any rate,
 it is useful to have such explicit statements.

\newpage

\begin{theorem} 
[Topological invariants/structures associated to M2-branes]
For the M2-brane in M-theory with oriented Riemannian 
worldvolume $M^3$ we have: 

\item {\bf (i)} For $M^3$ as framed manifold, the canonical framing provides a minimum for the classical action.

\item {\bf (ii)} Under the change of framing of the bounding 4-manifold $W^4$, the flux quantization 
condition of the C-field takes the form $2[F_4] + \lambda= 2a$ for $\phi \mapsto \phi + \sigma$ 
and $[F_4]+ \tfrac{1}{2}\lambda=a$ for $\phi \mapsto \phi + \rho$, where 
$\sigma$ and $\rho$ are the generators of $\pi_3({\rm SO}(4))=\Z \oplus \Z$.

\item {\bf (iii)} The M2-brane partition function is anomaly-free
if the framing of the worldvolume is even.

\item {\bf (iv)} The combination $[F_4] + \tfrac{1}{2}\lambda$ is an integral class if the Hopf invariant of the extended 
worldvolume of the M2-brane vanishes.

\item {\bf (v)} The dynamics of the C-field and its dual are encoded 
in the rational homotopy of the 4-sphere. 

\item {\bf (vi)} In the  formulation of the action of the M2-brane using the signature, the 
corner correction from the self-dual string is given by the Maslov index $\mu$
as $\exp (2\pi i \tfrac{1}{8}\mu)$. 

\item {\bf (vii)} The self-dual string as the boundary of the M2-brane on the M5-brane 
worldvolume defines
 a string theory in the sense of cobordism of manifolds with corners. 

%
%
%
\end{theorem}

\begin{theorem}
[Topological invariants/structures associated to M5-branes]
For the M5-brane in M-theory with oriented Riemannian 
worldvolume $M^6$ we have:

 {\bf (i)} The topological action for a framed M5-brane 
is given by the change of framing formula. The resulting action 
depends on the choice of framing. 

\item {\bf (ii)} The degree eight cohomology class, given by the lift of the electric current 
of the C-field and describing the 
dynamics on the M5-brane, is given by an element of Hopf invariant two.

\item {\bf (iii)} 
 The topological action of the M5-brane in eight
dimensions, both at the rational and integral levels, is given by (a
differential refinement of) the Hopf invariant. 


\item {\bf (iv)} For a Spin worldvolume $M^6$ and a normal bundle with a Membrane structure,
the vanishing of the Kervaire invariant implies that the effective action of the M5-brane 
is quadratic.

\item {\bf (v)} The contribution to the effective action due to the corner is given 
by the $f$-invariant.  

\item {\bf (vi)} In the signature formulation of the action of the M5-brane
in eight dimensions, the 
corner correction from the six-dimensional worldvolume $M^6$ is given 
by the Maslov index $\mu$
as $\exp (2\pi i \tfrac{1}{8}\mu)$. 

\item {\bf (vii)} The canonical $G_2$ structure minimizes the topological 
part of the M5-brane action, via the $\nu$-invariant.

\end{theorem}
%

The field strength $F_4$ and its dual satisfy quantization conditions 
characterized by the Pontrjagin classes of the M-theory target space $Y^{11}$
\cite{Flux} \cite{DMW} \cite{DFM}. 
The author has previously asked the question of 
whether these two fields can be viewed as components of 
one field strength which satisfies a general quantization condition, the components
of which yields the expected conditions.
It was also proposed that the total field strength might live in a generalized
cohomology theory (see \cite{tcu}). In this direction, building on Theorem {\bf 1(v)}, we have 

\paragraph{Proposal/conjecture.}
{\it The dynamics of the form fields in M-theory is 
described via the 
 spectrum ${\cal S}^4$ corresponding to the 4-sphere $S^4$. }

\vspace{2mm}
\noindent A discussion on justification of this statement, as well as Theorem {\bf 1(v)} 
is given in Section \ref{Sec coh}.
The paper is  divided into two main sections: 
Sec. \ref{Sec frame} on invariants associated with framings and Sec. \ref{Sec corner}
on those associated with corners. 
The M2-brane with its various notations of framings
are discussed in Sec. \ref{M2 frame}, where also the statements of Theorem {\bf 1({i--iii})}
are addressed. 
The M5-brane as a framed 
submanifold is described in Sec. \ref{M5 frame}, where Theorem {\bf 2(i)} is explained.  
Theorem {\bf 1(iv)} and Theorem {\bf 2(ii-iii)} are described in Sec. \ref{Rat}, Sec. \ref{Int}, 
and Sec. \ref{Z2}. The general discussion on relevance of corners is taken in Sec. \ref{M co}. 
The statements about the signature, i.e. Theorem {\bf 1(vi-vii)} and 
Theorem {\bf 2(vi)} are elaborated in Sec. \ref{Mas}. The remaining statements in Theorem 
{\bf 2}, i.e. parts {\bf (iv)} {\bf (v)}, and  {\bf (vii)} are explained in Sec. \ref{Ker}, 
Sec. \ref{Sec f} and Sec. \ref{Sec G2}, respectively. 
In the latter we also define a variant of the $\nu$-invariant arising from the one-loop
polynomial.

\medskip
We note that it has taken a while for this paper to appear in print. Meanwhile, the above proposal has 
been developed recently further \cite{FSS16a}\cite{cohomotopy},  leading to novel 
descriptions and connections to T-duality \cite{FSS16b}\cite{Loo}\cite{Higher-T}, 
twisted K-theory \cite{FSS16b}\cite{GaugeEnhancement}, and M-branes on orbifold singularities 
\cite{HSS}.

\section{M-branes as framed submanifolds and topological invariants}
\label{Sec frame}

  
  We start by outlining the various structures  which an
  M-brane can possess. 
  We will consider framings for both the M2-brane and the M5-brane. 
 Furthermore, the M2-brane will allow for more than one type/notion of 
  framing. 
Using \cite{KM}, we consider the set of framings $\varphi$, 
stable framings $\phi$, and  2-framings ${}^2\phi$,
which are compatible with a Spin structure $s$ on the worldvolume.



\begin{enumerate}

\item {\it Framing}: A framing $\varphi$ of 
the tangent bundle of the M2-brane worldvolume $M^3$, as
an oriented bundle, is a homotopy class
of sections of the associated frame bundle with structure group ${\rm SO}(3)$. 
Considering $M^3$ to be oriented or Spin, such a framing always exists. 
Denote the set of framings by $\cF$, and
 by $\cF_s$ the set of framings which are compatible with a Spin structure
$s$ on $M^3$. The latter is an affine space with translation group $\pi_3({\rm SO}(3))=\Z$
and is (non-canonically) isomorphic to $H^1(M^3; \Z_2) \oplus \Z$. 

\item {\it Stable framing}: 
The topological study of the M2-brane is often facilitated by extending to 
four dimensions. One can then consider structures on this coboundary,
whose tangent bundle is 
$\varepsilon^1 \oplus TM^3$, where $\varepsilon^1$
is an oriented line bundle.
A framing of this Whitney sum 
 is a stable framing $\phi$ of the worldvolume $M^3$. 
Denote the set of stable framings by $\F$ and the 
ones compatible with a Spin structure $s$ by $\F_s$. Similarly, 
$\F_s$ is an affine space with 
translation group $\pi_3({\rm SO}(4))=\Z \oplus \Z$  and is (non-canonically) isomorphic to 
$H^1(M^3;\Z) \oplus \Z \oplus \Z$.
This is analogous to the statement that a Spin${}^c$ structure on a manifold 
$X$ is equivalent to a Spin structure on $\varepsilon \oplus TX$
(see \cite{DMW-Spinc}). 
 Note that each framing $\varphi$ of the M2-brane worldvolume $M^3$ 
 can be identified with the stable framing  $\phi=\varphi \oplus \varphi_1$,
  where $\varphi_1$  is a framing of $\varepsilon^1$.

\item {\it 2-framing}: A 2-framing ${}^2\phi$ of the worldvolume $M^3$ 
is a homotopy class of 
trivializations of the twice the tangent bundle $2TM^3=TM^3 \oplus TM^3$.
This corresponds to the inclusion of structure groups 
${\rm SO}(3) \times {\rm SO}(3) \hookrightarrow {\rm SO}(6)$.
The set of 2-framings ${}^2F$ is an affine space with translation group
$\pi_3({\rm SO}(6))\cong\Z$.  
This is related to String structures 
\cite{BN} and applied to describe the C-field and 
the M2-brane in \cite{tcu} \cite{String}. The main connection is via the quantization condition 
of the C-field \cite{Flux}, which implies 
 that $C_3$ is essentially (but not literally) the difference of two Chern-Simons
forms; see \cite{tcu}.

\end{enumerate}

We have seen that a framing is a trivialization of (some variant of)
\footnote{This is strictly speaking a trivialization  of a class in real K-theory $\widetilde{KO}$. 
However, by what seems like an accepted abuse, we will take this to be a trivialization of a 
particular bundle when no confusion arises.}
the tangent bundle of the worldvolume. This is the {\it tangential framing}. 
There is another trivialization that is also important to the M-branes, namely the
{\it normal framing}.

  \paragraph{The M-branes as framed submanifolds.}
 A normally framed $m$-submanifold of an $n$-dimensional 
 manifold $N$ is a submanifold $M$ with a given 
 framing 
 $
 f: {\cal N}_M N\cong M \times \R^{n-m}
 $
 of the normal bundle ${\cal N}_M$. 
Let us consider the situation in M-theory on $Y^{11}$. 
For a normally framed M2-brane with worldvolume 
$M^3$, we have 
$f_3: {\cal N}_{M^3}  Y^{11} \cong M^3 \times \R^8$.  
The main examples of framed three-dimensional 
worldvolumes are 3-tori, 3-spheres and their 
quotients, such as lens spaces.  
 Similarly, for a normally framed M5-brane with worldvolume 
 $M^6$, we have
 $f_6: {\cal N}_{M^6}  Y^{11} \cong M^6\times \R^5$.  
Main examples of such M5-brane worldvolumes are  
the 6-sphere, the products $G_1 \times G_2$ of any of the
Lie groups
${\rm SO}(3)$ and ${\rm Sp}(1)={\rm SU}(2)={\rm Spin}(3)$ 
and their quotients by finite groups (see \cite{F}).

\paragraph{The M-branes when $Y^{11}$ is a parallelizable manifold.}
We have previously discussed geometric and topological consequences
of having the target spacetime $Y^{11}$ to be a framed or parallelizable 
manifold  \cite{String} \cite{F}. We now extend an aspect of the 
 description in the
presence of 
M-branes. 
Consider $Y^{11}$ as a $\pi$-manifold and take $M^m$ to be the worldvolume of 
the M2-brane ($m=3$) or of the M5-brane ($m=6$). 
Since $TM^m \oplus {\cal N} M^m = T_M Y^{11}$ then 
$
TM^m \oplus {\cal N} M^m \oplus \varepsilon^k= T_M Y^{11} \oplus \varepsilon^k = \varepsilon^{11+k} 
$
so that the normal bundle ${\cal N} M^m$ is stably trivial if and only if the worldvolume
$M^m$ is a framed manifold. 
Next we consider the disk bundle, needed for the description of tubular neighborhood
for the M5-brane
 (see \cite{DFM}\cite{boundary}). 
Let $X$ be the total space of the disk bundle over the worldvolume 
$M^m$, which we take to be a framed manifold, associated to a vector
bundle $\eta$. We identify $M^m$ with the zero section of 
$X$. Then $TX$ is stably trivial if and only if $TX|_M$ is stably trivial. Since 
$TX|_M=TM^m \oplus {\cal N} M^m= TM^m \oplus \eta$, and since
$TM^m$ is stably trivial, then $X$ is parallelizable if and only if 
$\eta$ is stably trivial.

\subsection{The framing and stable framing on the M2-brane}
\label{M2 frame}

We consider the M2-brane worldvolume $M^3$  
with 
a Spin structure.
$M^3$ as a framed manifold is discussed extensively in 
\cite{tcu}. 

 \paragraph{Chern-Simons theory and $p_1$-structure.}
A closely related structure to both a 2-framing and a String structure is the 
notion of {\it $p_1$-structure} \cite{BHMV}, or {\it rigging} \cite{Se}, arising 
in the context of topological field theory, especially Chern-Simons theory. 
Analyzing the worldvolume anomalies of the M2-brane leads to the 
quantization condition on the C-field \cite{Flux}, given by 
  $[F_4]- \frac{1}{2}\lambda=a$, where $F_4$ is the field strength 
-- curvature, in the description as a 2-gerbe/3-circle bundle   \cite{FSS1} \cite{FSS2} -- of
$C_3$, $\lambda=\tfrac{1}{2}p_1$ is the first Spin characteristic class 
of the Spin bundle, and $a$ is the degree four characteristic class of an $E_8$ bundle. 
Therefore, a main aspect of the C-field is captured by Chern-Simons theory. 
The path integral in this theory depends on the orientation as well as on the 
$p_1$-structure. Such a structure 
arises since the invariants 
considered turn out to have a {\it framing anomaly} \cite{Jones}, i.e. the 
invariants themselves depend on the $p_1$-structure. This has an interpretation in terms of 
String structures on the M-branes \cite{tcu}.

\medskip
 Let $X={\rm BO}\langle p_1 \rangle$ be the homotopy fiber of the map 
$p_1: BO \to K(\Z, 4)$ corresponding to 
the first Pontrjagin class of the universal stable bundle $\gamma$ 
over the classifying space $BO$. Let $\gamma_X$ be the pullback of 
$\gamma$ over $X$. A $p_1$-structure on the worldvolume $M^3$ is a fiber
map from the stable tangent bundle $TM^3$ of $M^3$ to $\gamma_X$. 
That is, there is the following lifting diagram
\(
\xymatrix{
&& X={\rm BO}\langle p_1 \rangle
\ar[d]
&&
\\
M^3
\ar[urr]
\ar[rr]
&& 
{\rm BO}
\ar[r]^-{p_1}
&
K(\Z, 4)\;.
}
\)

The Spin/String version of this 
 construction is explained in our context in 
\cite{tcu}. The study of the M2-brane anomalies require the 
extension to four dimensions.   The above discussion 
shows that this is not automatic in the oriented case as there is an 
obstruction; the cobordism group $\Omega_3^{p_1}$ of oriented 3-manifolds
with $p_1$-structure, is isomorphic to $\Z/ 3\Z$, the isomorphism 
being induced by the invariant $h$, described shortly below. 
The description of the effective action in M-theory via 
the signature of oriented  manifolds in place of Dirac operators of Spin manifolds 
is given in \cite{sig}. Strictly speaking, anomalies suggest the use of 
Spin structure, but one can ignore this structure for other purposes 
and consider the expression $4[F_4]- p_1=4a$. Note that the quantization
condition does not necessarily assume $\lambda$ to be 
divisible by two, and in many cases it is not \cite{tcu} \cite{top2}, 
so that the expression in this case is $2[F_4] -\lambda=2a$. 
We will consider this again, more appropriately  
in the case of the M5-brane, where 
an oriented worldvolume is more justifiable.

\medskip
The relevant group for a String structure  is the stable homotopy/cobordism group 
$\pi_3^s \cong \Omega_3^{\rm fr} \cong \Omega_3^{\rm String} \cong \Z_{24}$,
which can be seen from the worldvolume \cite{tcu}. 
 We now describe how this can be seen straightforwardly via the embedding to 
 11-dimensional spacetime.

\paragraph{The group $\Z_{24}$ from eleven dimensions.}
Consider the M2-brane worldvolume $M^3$ embedded in the ambient 11-dimensional 
spacetime via  $i: M^3 \hookrightarrow Y^{11}$. The simplest situation is a product
 $Y^{11}\cong M^3 \times \R^8$.
Write $M^3_+$ for the disjoint union of $M^3$ with a disjoint base-point. Then there is 
a canonical homeomorphism ${\cal T}(M^3 \times \R^8)\cong \Sigma^8(M^3_+)$ between the 
Thom space of the trivial $8$-dimensional vector bundle and the $8$-fold suspension
of $M^3_+$: 
$(S^8 \times M^3_+)/(S^8 \vee M^3_+)=S^8 \vee M^3_+$.
Now suppose that $M^3$ is given with a choice of trivialization of the normal 
bundle ${\cal N}(M^3,i)$. This choice is a choice of homeomorphism 
${\cal T}({\cal N}(M^3,i))\cong \Sigma^8M^3_+$, which is a framing of $(M^3, i)$. 
Now identify the $11$-dimensional sphere $S^{11}$ with the 
one-point compactification $\R^{11}\cup \{\infty \}$. 
The Pontrjagin-Thom construction is the map
$S^{11} \to {\cal T}({\cal N}(M^3,i))$
given by collapsing the complement of the interior of the unit disk
bundle $\mathbb{D}({\cal N}(M^3,i))$ to the point corresponding to $S({\cal N}(M^3,i))$
and by mapping each point of 
$\mathbb{D}({\cal N}(M^3,i))$ to itself. 
Identify the $8$-dimensional sphere with the $8$-fold suspension 
$\Sigma^8S^0$ of the zero-dimensional sphere (i.e. two points,
one of which is the base-point). The map which collapses $M^3$ 
to the non-basepoint yields a base-point preserving 
map $\Sigma^8(M^3_+) \to S^8$ whose homotopy class defines
an element of $\pi_{11}(S^8)\cong \Z_{24}$.

 \paragraph{Extending the $C$-field framing to the bounding 4-manifold.} 
 The extension of the C-field to the bounding 4-manifold $W^4$ in the Spin case is independent 
 of the choice of this 4-manifold \cite{Eff} \cite{Flux}. 
 A similar observation holds in the case of String structures \cite{tcu}, where 
 the relation to gerbes is highlighted. We now provide
 a description in terms of (variants of) framings. 
  Since $H^4(W^4)=0$, the Pontrjagin class $p_1(E)$ for 
 all real vector bundles over $W^4$ vanishes. Therefore, we should instead 
 consider relative characteristic classes, as done in \cite{tcu}. The relative Pontrjagin number 
 is defined as the pairing of the relative Pontrjagin class with the 
 fundamental class
 $
 p_1(W^4, \varphi)=p_1(TW^4, \varphi)[W^4, M^3]
 $
 associated to any given tangential framing $\varphi$ over $M^3$. Thus
 $p_1(W^4, \varphi)$ is an integer invariant which measures the obstruction to 
 extending $\varphi$ to a framing of the tangent bundle of $W^4$.
 The 
 Hirzebruch defect of a framing $\varphi$ of the closed oriented 3-manifold 
 $M^3$ is 
 \(
 h(\varphi)= p_1(W^4, \varphi) - 3 {\rm sign} (W^4)\;.
 \)
 It follows from the Novikov additivity of the
 signature and the signature formula for closed manifolds that 
$h(\varphi)$ is independent of the choice of $W^4$. 
This is a multiple of the Atiyah-Patodi-Singer eta-invariant
of the (odd) signature operator is given by 
 $
 \eta (M^3):= \int_{W^4} L - {\rm sign} (W^4)
 $,
 where 
 the metric on some collar neighborhood of $\partial W^4$ is 
 isometric to the product metric on $M^3 \times [0, \epsilon )$, and
 $L$ is the Hirzebruch $L$-class as a degree four class $\frac{1}{3}p_1(W^4)$
on $W^4$. 

\vspace{3mm}
The relation between the eta-invariant 
$\eta (M^3)$ and the Chern-Simons invariant $CS(M^3)$ 
for a compact 3-manifold $M^3$ is 
$
3 \eta (M^3) \equiv 2CS(M^3) + \tau$ (mod $\Z$),
 where $\tau$ is the number of 2-primary summands of the 
first homology group $H_1(M^3; \Z)$. Thus the eta-invariant 
completely determines the Chern-Simons invariant once the 
homology of the worldvolume $M^3$ is known. 
 When $\tau=0$ the Hirzebruch defect $h(\varphi)$
 is given by a multiple of the Chern-Simons invariant 
 \(
 2CS(M^3) \equiv h(\varphi)
 \mod \Z\;.
 \)
 Part of the M2-brane partition function is captured by the holonomy
 of the C-field, which is essentially given by Chern-Simons theory. 
 The holonomy is given by (see \cite{DFM})
 \(
 \chi (C_3)= \exp \left[ 
 2\pi i \left( 
 CS(A) - \tfrac{1}{2}CS(g) + c
 \right)
 \right]\;,
 \)
 where $CS(A)$ is the Chern-Simons invariant of a connection $A$ of 
 an $E_8$ bundle, $CS(g)$ is the `gravitational' Chern-Simons invariant 
 corresponding to a metric $g$ on $M^3$ (or to the corresponding 
 Levi-Civita connection on $M^3$), and $c$ 
 is a constant background 3-form.

\paragraph{The M2-brane effective action and canonical tangential framings.}
We will see that there is a framing of the M2-brane worldvolume $M^3$ 
which is, in a sense, preferred. In \cite{tcu} \cite{String} the canonical String structure 
of \cite{Red} \cite{BN} was highlighted as the one preferred by 
the M2-brane. Here we elaborate, highlighting some dynamical aspects. 
A Spin structure $s$ over $M^3$ can be considered as a framing of $TM^3$
over the 2-skeleton of $M^3$, and 
consider the set $\cF_s$ of framings of $M^3$ which are compatible with 
$s$. From obstruction theory, 
the difference between two such framings is specified by an element 
of the cohomology group $H^3(M^3; \pi_3({\rm SO}(3)))=\Z$.
 A framing $\varphi$ of the worldvolume $M^3$
is {\it canonical} for the Spin structure $s$ if it is compatible with $s$, and 
$|h(\varphi)| \leq |h(\varphi')|$ for all other framings $\varphi'$ which are compatible
with the Spin structure $s$. This means that $\varphi$ is a minimum for the
invariant $|h|$ on $\cF_s$ \cite{KM}. The discussion in \cite{tcu} \cite{String} shows that
this invariant is  the effective action of the M2-brane. Therefore,  we have
a minimization of this effective action when we choose the canonical framing. 
This is a classical statement. A corresponding quantum (or at least semi-classical) 
statement might involve anomalies;
this is what we consider next, but using stable framings instead.   
Dependence on Spin structure is considered in \cite{DMW-Spinc}
 while dependence of the M2-brane on String structures is
 discussed in \cite{tcu}. We now consider dependence on framing.

%
%
%
%
%



\paragraph{A possible M2-brane anomaly via canonical stable framings.}
We will discuss this from the point of view of both the C-field $C_3$ and 
its field strength $F_4$. The first involves dealing with 
secondary classes such as the eta-invariant and the Chern-Simons invariant,
while the second involves the primary classes, mainly the first Pontrjagin class. 
First, concentrating on the latter, 
we consider the effect of change of `usual' framings and show how that is
anomaly-free.  
Now we extend to the bounding 4-manifold and 
ask whether there is a dependence on framing  or whether there is 
a corresponding potential anomaly. 
For two framings $\varphi_1$ and $\varphi_2$ of $M^3$, 
 the Pontrjagin class
 has the property 
 $
 p_1(W^4, \varphi_1 \oplus \varphi_2)= p_1(W^4, \varphi_1) + 
 p_1(W^4, \varphi_2)$.
 Correspondingly, for any two framings $\varphi_1$ and $\varphi_2$ of $M^3$, the
 Hirzebruch defect has the 
the additivity property
 $
h(\varphi_1 \oplus \varphi_2)= h(\varphi_1) + h(\varphi_2)$,
 which induces a similar property for the Chern-Simons 
 invariant. The effect of such a change on the 
 holonomy of the C-field is multiplicative and there is no 
 potential anomaly. 
 
\vspace{3mm}
Next we look at stable framings.
Let $\phi$ be a stable framing of $M^3=\partial W^4$. 
Then, as shown in \cite{KM}, the effect of translation by 
the generators 
\footnote{
Explicit generators $\rho$ and $\sigma$ for $\pi_3({\rm SO}(4))=\Z \oplus \Z$
 can be found in \cite{Steenrod} (see also \cite{KM}). 
View $S^3$ as the unit sphere in the quaternions $\H$, oriented by the 
ordered basis $1, i, j, k$, and view SO(4) as the rotation group of $\H$. 
Then, for $q\in \H$ and $x \in S^3$, 
the maps $\rho$ and $\sigma: S^3 \to$ SO(4) defined by 
$\rho(q)x=qxq^{-1}$ and $\sigma(q) x=qx$ represent generators of 
$\pi_3({\rm SO(4)})$. 
By restricting to ${\rm Im}\H$ of pure (imaginary) quaternions, $\rho$ also
represents a generator of $\pi_3({\rm SO}(3))$, and these two $\rho$'s
correspond under the natural map $\pi_3({\rm SO}(3)) \to \pi_3({\rm SO}(4))$
induced by the inclusion SO(3) $\hookrightarrow$ SO(4).}
$\rho$ and $\sigma$ of $\pi_3({\rm SO}(4))=\Z \oplus \Z$  
is given by 
\(
p_1(W^4, \phi + \rho)= p_1(W^4, \phi)+ 4\;, \qquad \qquad 
p_1(W^4, \phi + \sigma)= p_1(W^4, \phi) +2\;.
\label{pont}
\)
In the flux quantization condition on the C-field, 
$[F_4] + \tfrac{1}{2}\lambda =a \in H^4(Y^{11}; \Z)$, 
the 3-sphere $S^3$ was used as a representative case for the 
M2-brane worldvolume $M^3$. This is further considered and generalized 
in \cite{tcu} from the point of view of framed 3-manifolds.  In the general
case, one has that the flux quantization holds on $W^4$ by considering the 
extension of spacetime $Y^{11}$ to bounding $Z^{12}$ in such a way that
the extra directions in both cases can be identified. Then 
the transformations \eqref{pont} suggest that the change of stable framing by 
$\rho$ is allowed, while that by $\sigma$ leads to a potential anomaly
since $\tfrac{1}{2}\lambda$ is not defined. However, this is not a drastic anomaly 
since one can modify what one means by the quantization condition in such
situations, i.e. multiply through by 2. 
Nevertheless, this shows that at least the transformation $\sigma$ 
requires care.  
Note that  a stable framing $\phi$ on $M^3$ which is compatible with 
a Spin structure $s$ on $M^3$
extends to a framing of a compact  4-manifold $W^4$ bounding 
$M^3$, then $p_1(W^4, \phi)=0$ \cite{KM}.
The consequence of divisibility of $\lambda$ by 2 is highlighted in 
\cite{top1} leading to the notion of a Membrane structure, that is 
having $w_4=0$.

 \medskip
 Next we consider one effect of the change of framing. 
 Since the C-field is essentially a Chern-Simons form 
 (in the sense explained above), we can rely on the effect of change of 
 framing on the Chern-Simons form itself. 
 Under a change of framing $\varphi \mapsto \varphi + s$, 
 the gravitational Chern-Simons form transforms as 
 $CS(\omega) \mapsto CS(\omega) + 2\pi s$, and the 
 partition function transforms as $Z \mapsto Z\cdot \exp \left( 
 2\pi i \cdot \frac{s}{24}
 \right)$ \cite{Jones} \cite{JM}.
 This is described using String cobordism in \cite{tcu}.
 Taking $C_3 = CS(A) - \tfrac{1}{2}CS(\omega)$, we get 
 $C_3 \mapsto C_3 - \pi s$ and $e^{iC_3} \mapsto e^{iC_3} \cdot e^{-i\pi s}$. 
This shows that, 
 unless the other factor in the partition function, namely the 
 Pfaffian ${\rm Pfaf}(D)$ of the Dirac operator $D$, cancels this 
 anomalous factor, we see that $s$ has to be even.

\paragraph{Restriction to the heterotic boundary.} 
We now consider the M2-brane with boundary $\partial M^3$, where this boundary does
not lie on the M5-brane in eleven dimensions, but rather restricts to the 
heterotic string on the boundary 
of the ambient 11-manifold. 
Let $Y^{11}$ be a manifold with boundary $X^{10}=\partial Y^{11}$
and let ${\cal L}$ be a subbundle of $T_X Y^{11}$ spanned by a vector 
field $v$ pointing inside $Y^{11}$. Then 
$T_X Y^{11}= TX^{10} \oplus {\cal L}$.
Now we take $M^3 \subset Y^{11}$ to be a neat submanifold, 
that is, the intersection of the tubular neighborhood 
of the M2-brane worldvolume $M^3$ with $X^{10}$ is a tubular neighborhood of 
the string worldsheet $\Sigma=\partial M^3$ in $X^{10}$. 
Then 
we can assume that along $Y^{11}$ the vector $v$ points inside $M^3$. 
Then the tangent bundle to the M2-brane worldvolume splits as 
$T_\Sigma M^3= T\Sigma \oplus {\cal L}$, where 
$\Sigma$ is the heterotic string on the boundary. 
There is a natural identification 
$
T_\Sigma Y^{11}/T_\Sigma M^3= T X^{10}/T\Sigma
$
of bundles restricted to the string worldsheet $\Sigma$. 
That is, the normal bundle of the M2-brane worldvolume $M^3$ restricted to 
its string boundary worldsheet $\Sigma$ can be identified with the normal 
bundle of the string worldsheet $\Sigma$ in the heterotic target 
space $X^{10}=\partial Y^{11}$.

\subsection{The M5-brane as a framed submanifold}
\label{M5 frame}

We now turn to the framing for the M5-brane.  
The discussion will 
rely not only on that worldvolume but also on the interplay with the normal bundle 
in eleven-dimensional spacetime. 
Other notions of framings has been used before, but they are different from 
the (standard) one we use here. For example, 
\cite{BCR} consider the M5-brane to be framed, whereby 
the normal bundle splits off a one-dimensional 
trivial summand so that the structure group reduces 
from SO(5) to SO(4), as in a case in \cite{Eff}. 
We will consider more general and standard decompositions, 
where the normal bundle completely decomposed into  a Whitney sum of trivial line bundles. 

\medskip
The M5-brane need not necessarily be Spin, but can be taken to 
be oriented. Therefore, the discussion in 
\cite{tcu2} \cite{top1} can be adapted to this case, resulting in 
twisted versions of $p_1$-structures, rather than twisted versions of 
String and String${}^c$ structures, on the worldvolume. 
Note that such a  situation can be interpreted also in 
terms of elliptic cohomology \cite{KS1}.
Therefore, instead of conditions of the form $\tfrac{1}{2}p_1 + \alpha=0$
for various degree 4 classes $\alpha$, 
we will have a condition of the form $p_1 + \alpha=0$.  
In the spirit of the approach of \cite{Wa} \cite{SSS3} 
\cite{tcu} \cite{top1} \cite{top2}, we will interpret this latter 
condition as a twisted structure:

\begin{definition}
An $\alpha$-twisted $p_1$-structure on a brane $\iota: M \to Y$ 
with a Riemannian structure classifying map $f: M \to BO$, is a 4-cocycle
$\alpha: Y \to K(\Z,4)$ 
and a homotopy $\eta$ in the diagram 
$$
    \raisebox{20pt}{
    \xymatrix{
       M
       \ar[rr]^-{f}_>{\ }="s"
       \ar[d]_\iota
       &&
       B \mathrm{O}(n)
       \ar[d]^-{p_1}
       \\
       Y
       \ar[rr]_\alpha^<{\ }="t"
       &&
       K(\mathbb{Z},4)\;.
       \ar@{=>}^\eta "s"; "t"
    }
    }
  $$
  \label{def}
\end{definition}
The obstruction is then $p_1(M) + [\alpha]=0\in H^4(M; \Z)$. 
As in the twisted String case, the set of such structures will be a torsor for $H^3(M; \Z)$.

\paragraph{Framing of the M5-brane as a submanifold.}
A {framing} of the M5-brane worldvolume as 
a submanifold $M^6 \subset Y^{11}$ 
of dimension $6$ is the assignment 
$\xi$ of $5$ linearly independent vectors 
$
(\xi^1(x), \cdots, \xi^5(x)) 
$ 
in $T_xY^{11}$ that are normal to $M^6$. 
The pair $(M^6, \xi)$ is then a {Pontrjagin framed manifold}. 
A framing may be thought of an an isomorphism between the normal bundle
of the embedding and the Whitney sum of five copies of the 
(unique) oriented line bundle over $M^6$. The importance of this arises from the
fact that the spinors on the M5-brane take values in the Spin bundle of the normal 
bundle. With the latter being trivial, the former is also trivial and hence 
admits a maximal number of sections; that is, such a configuration 
is maximally supersymmetric. 

\vspace{3mm}
Let $i: M^6 \to \R^{6+r}$ be an embedding in Euclidean space for large enough $r$. 
The normal bundle ${\cal N} (M^6, i)$ of $i$ is the quotient of the pullback 
of the tangent bundle of $\R^{6+r}$ by the sub-bundle given by the tangent
bundle of $M^6$, ${\cal N}(M^6, i)=i^*T\R^{6+r}/TM^6$, so that ${\cal N}(M^6, i)$ is an 
$r$-dimensional real vector bundle over $M^6$. 
If we give $T\R^{6+r}=\R^{6+r} \times \R^{6+r}$ 
the Riemannian metric obtained from the usual inner product in Euclidean space, 
 ${\cal N}(M^6, i)$ may be identified with the orthogonal 
complement of $TM^6$ in $\i^*T\R^{6+r}$. That is, the fiber at $z\in M^6$ 
may be identified with the subspace of vectors $(z,x)\in \R^{6+r} \times \R^{6+r}$
such that $x$ is orthogonal to $i_*(TM^6)_z$, where $i_*$ is the induced embedding of 
$TM^6$ into $T\R^{6+r}$. 
If $M^6$ admits an embedding with trivial normal bundle, we say that $M^6$ has 
a {stably trivial normal bundle}. 
Different embeddings lead to different normal bundles, but stably they all
coincide with the {stable normal bundle} which is classified by the 
Gauss map ${\cal N}(M^6) \to BO=\lim_k BO(k)$. 
If $r$ is 
sufficiently large 
and $i_1, i_2: M^6 \to \R^{6+r}$ are
two embeddings, then ${\cal N} (M^6, i_1)$ 
is trivial (i.e. ${\cal N}(M^6, i_1) \cong M \times \R^r$) if and only if 
${\cal N} (M^6, i_2)$ is trivial.



\vspace{3mm}
In general, $r$ has to be more than twice the dimension of the submanifold, that 
is $r> 12$ in the case of the fivebrane. However, 
one can take the topology of $M^6$ to be such that
this still works for lower codimension. Therefore, we will assume that we are in such a setting 
and will concentrate on the 
case when we have a physical embedding in the 11-dimensional
M-theory target space $Y^{11}$. 
Note that for the case of the M2-brane, this is automatically satisfied 
because of the already large codimension. 
Further discussion
 related to embeddings of the M-branes can be found in \cite{Target}.

\paragraph{The M5-brane via the Pontrjagin-Thom construction.}
 Similarly to the M2-brane, the M5-brane can be described straightforwardly using the 
Pontrjagin-Thom construction. 
There is 
a canonical homeomorphism ${\cal T}(M^6 \times \R^5)\cong \Sigma^5(M^6_+)$ between the 
Thom space of the trivial $5$-dimensional vector bundle and the $5$-fold suspension
of $M^6_+$.
The map which collapses $M^6$ 
to the non-basepoint yields a base-point preserving 
map $\Sigma^5(M^6_+) \to S^5=\Sigma^5 S^0$ 
whose homotopy class defines
an element of $\pi_{11}(S^5)\cong \Z_2$. 
Therefore, instead of $\Z_{24}$ for the M2-brane we have $\Z_2$
for the M5-brane. 
This is appropriate for considering the relation to the Kervaire 
invariant rather than to String structures, although there are 
relations of the M5-brane to the latter  manifested in a different 
way (see \cite{tcu2} \cite{FSS1} \cite{FSS2}).
We will now highlight a consequence of that. 

\paragraph{The String $E_8$ bundle and the octic invariant.}
We now consider the effect of having an $E_8$ bundle in six dimensions,
as suggested by the presence of the C-field. 
We take $M^{6}$ to be a compact framed 6-manifold without a boundary
admitting a String structure (see \cite{tcu2}  \cite{FSS1} for
 justification). 
More precisely, we take a String $E_8$ bundle on $M^6$, that is, a bundle of the 
3-connected cover $E_8\langle 3 \rangle$ of Lie group $E_8$. 
Thus there is no transgressed degree three class that would characterize the 
bundle; instead, there will be a class in dimension 15 corresponding to 
the generator of $H^{16}(BE_8;\Z)\cong \Z$. 
This octic invariant has an explicit description (see \cite{CP}), 
to which we now propose  a connection. 
Consider a map $f: M^{6} \to BE_8\langle 4 \rangle$ to the classifying space 
$BE_8\langle 4 \rangle$ of $E_8\langle 3 \rangle$ String bundles
on $M^6$. 
For such a map, the Pontrjagin-Thom construction yields the isomorphism
$
S^{6+5}\longrightarrow {\cal T}({\cal N}(M^6,i))\cong \Sigma^5M^{6}_+ \longrightarrow
\Sigma^5 {BE_8\langle 4 \rangle}_+$,
whose homotopy class defines an element of $\pi_{6+5}(\Sigma^5({BE_8\langle 4 \rangle}_+))$. 
Now, to `lowest order',
\footnote{One usually has the approximation $E_8 \sim K(\Z,3)$, but then considering 3-connected covers kills 
topology in dimension 3, and if one considers the next nontrivial class, this would be in 
dimension 15, though this resulting structure will be topologically trivial.
}
 the homotopy type of 
$BE_8\langle 4 \rangle$ can be `approximated' by that of the Eilenberg-MacLane 
space $K(\Z, 16)$.
Then, taking $M^6$ to be framed and embedded in $\R^{11}$, 
we get a generator of $\pi_{11}(\Sigma^5(K(\Z,16))$. 
This is $\pi_{11}(K(\Z, 11))\cong \Z$ and corresponds to the 
octic invariant of $E_8$. Therefore, we can detect the generator of 
$\pi_{15}(E_8)$  or of $H^{15}(E_8; \Z)$, albeit rather indirectly and 
essentially in the topologically trivial sector.

%
%
%
%
%
%
%

\paragraph{Framed bordism of M-branes and corresponding  invariants.}
Two framed $n$-manifolds $(N_i, f_i)$ are bordant if there is a framed
 $(n+1)$-manifold $(B, g)$ with $\partial B=\{0\} \times N_0 \cup \{1\}\times N_1$,
 $\partial g=f_0 \cup f_1$. 
 The bordism classes $[N, f]$ of framed submanifolds form an abelian 
 group $\Omega_n^{\rm fr} N$.
For M2 and M5 branes with trivial normal bundles embedded in
spacetime $Y^{11}$, we consider the M-branes as boundaries 
or as admitting boundaries. 
 For example, we can use the above bordism picture to 
 describe framed open M2-branes 
 with boundaries on M5-branes \cite{St} \cite{To}
 or framed M5-branes with boundaries on M9-branes
 \cite{BGT}. 
This is discussed in \cite{tcu} for the case of the M2-brane.   
Certain expressions such as the effective action or the 
partition function sometimes are themselves topological invariants, 
or depend on topological invariants;
for instance, framed cobordism invariants. This implies that theses
expressions will have the same form or value for every element 
(i.e. space/worldvolume) in the cobordism class. Therefore, it is 
enough for this purpose to evaluate the expression, i.e., the 
effective action or the partition function, on generators. 
For the case of the M2-brane, the framed cobordism group is 
$\Omega_3^{\rm fr} \cong \Z_{24}$, generated by $S^3={\rm SU}(2)$ with the Lie group framing. 
For the M5-brane the corresponding cobordism group is
 $\Omega_6^{\rm fr}\cong \Z_2$, generated by $S^3 \times S^3={\rm SU}(2) \times {\rm SU}(2)$ 
 with the product Lie group framing. 

%
%

%
%
%
%
%
%
%
%
%


\paragraph{The topological pairing and the action in the presence of a framing.}
It is useful to formulate the M5-brane 
theory in seven dimensions 
\cite{FSS1} and consider an appropriate restriction to six dimensions
\cite{FSS1} \cite{FSS2} \cite{cup} \cite{up}. 
We now consider the effect of framing on the topological part
of the action, making use of results of Lannes \cite{La} \cite{Mi}.
We start with the M5-brane worldvolume $M^6=\partial N^7$
as a framed manifold and study mod 2 torsion 
cohomology classes of middle 
degree $H^3(M^6; \Z_2)$, corresponding to a lift of the class of the 
worldvolume gerbe.  
\footnote{The arguments we make here can be extended to differential cohomology. 
While this is interesting, we will not pursue it here as we are interested in topological
invariants. See also the end of Section \ref{Rat}.}
We take the inclusion of the boundary 
$i: M^6 \hookrightarrow N^7$, and discuss to which extent the gerbe 
on $M^6$ arises from the one on the extension $N^7$
when studying the effective action.
The self-duality of the exact sequence 
\(
\xymatrix{
H^3(N^7; \Z_2)
\ar[r]^-{i^*}
&
H^3(M^6; \Z_2)
\ar[r]^-{\delta}
&
H^4(N^7, M^6; \Z_2)
\ar[r]
&
H^4(N^7; \Z_2)
}
\label{SD}
\)
implies that ${\cal I}:={\rm Im}(i^*:H^3(N^7; \Z_2) \to H^3(M^6; \Z_2))$
is a Lagrangian subspace in the inner product space 
$E:=H^3(M^6; \Z_2)$. 
That is, ${\cal I}={\cal I}^\perp$ with respect to the symplectic form on $E$.

\vspace{3mm}
An element $x\in H^3(M^6;\Z_2)$ can be thought of 
as a homotopy class of maps of the one-point compactification of the 
worldvolume $M^6_+ \to K(\Z_2, 3)$, and so 
determines an element 
$
S^6 \to \Sigma^\infty M^6_+ \to \Sigma^\infty K(\Z_2, 3)
$
of the stable homotopy group $\pi_6(K(\Z_2, 3))$. This 
group is of order 2, so that the framing $f$ determines 
a map 
\(
q_f: H^3(M^6;\Z_2) \to \Z_2\;,
\)
i.e. 
$E$ can be equipped with this quadratic form.
If $N^7$ admits a framing extending that of $M^6$, then the
quadratic form is trivial on ${\cal I}$, and so the Witt class of the 
quadratic form is a framed cobordism invariant. 
In this case one can characterize the elements 
$u \in E$ such that $q(x)=u \cdot x$ for $x\in {\cal  I}$, in terms of the 
relative Wu class $v_4(\nu, f)\in H^4(N^7, M^6; \Z_2)$.
This class restricts on $N^7$ to $v_4(\nu)\in H^4(N^7; \Z)$, 
due to dimension.
Let $u \in H^3(M^6; \Z_2)$ be the class such that 
$\delta u =v_4(\nu, f)\in H^4(N^7, M^6; \Z_2)$. This is well-defined
modulo ${\cal I}$, and  
$q(x)=x\cdot u$ for any $x\in {\cal I}$. 
Take $x=i^*y$ for $y\in H^3(N^7; \Z_2)$. By self-duality 
of the sequence \eqref{SD}, this expression can be rewritten as
\(
q(i^*y)=i^*y \cup u=y\cup \delta u=y\cup v_4(\nu, f)\;.
\)


\paragraph{Effect of change of framing on the quadratic form.}
The intersection pairing determined by the framing 
leads to a quadratic refinement in the sense of 
Browder-Brown \cite{BP} \cite{Generalizations}. 
Now we consider the effect of change of framing via gauge transformations.
That is, we consider two framings $f_1, f_2$  differing by a gauge transformation 
$g: M^6 \to O(6)$. We can also consider this in the stable range, 
i.e. replace $O(6)$ by $O$.
Then one has the change of framing formula \cite{JR}
$q_{f_1}(x) + q_{f_2}(x)= x \cdot g^* z$, where 
$z$ is the degree three class in the diagram 
\(
\xymatrix{
O \ar[d] \ar[rr]^-{z} && K(\Z_2, 3)\ar[d]\\
EO \ar[d]  \ar[rr] && BO\langle v_4 \rangle\ar[d] \\
BO \ar[rr]^-{=} && BO\;.
}
\label{x3}
\)
The gauge group of smooth maps from $M^6$ to $O$ 
acts transitively on framings (with respect to this embedding), and 
the corresponding group of maps, the real 
K-group $KO^{-1}(M^6)=[M^6, O]$,
acts transitively on the set of stable framings. 
This is depicted in the following diagram
\(
\xymatrix{
&& EO \ar[d]^\pi && O
 \ar[ll]
\ar[rr]
&&
K(\Z_2, 3)
\\
M^6 
\ar@{..>}[urr]^f
\ar[rr]
&&
BO
\ar[rrrr]^{v_4}
\ar@{..>}[urr]_\Omega
&&
&&
K(\Z_2, 4)\;.
\ar@{..>}[u]^\Omega
}
\)
For $g$ a gauge transformation, Brown's theorem \cite{Generalizations} implies the transformation law
\(
q_{gf}(x)=q_f(x) + \langle
x\cup g^* \overline{v}_{4}, [M^6]
\rangle\;,
\)
where $\overline{v}_{4}=z$ denotes the image in $H^3(O; \Z_2)$ of the Wu class 
$v_{4}$ under the map 
$
\omega: \Sigma O \to BO
$, which is 
adjoint to the equivalence map $O \to \Omega BO$. 
Therefore, this gives a formula for the action in 
six dimensions. Furthermore, the action in six dimensions depends on the choice of framing, since 
$\frac{6}{2}-1$ is a power
\footnote{This criterion works in higher dimensions as well.}
 of 2. 

\paragraph{Example.} 
Consider the case when the worldvolume is a product $M^6=M_1^3 \times M_2^3$
of two 3-dimensional framed manifolds. 
Consider two gauge transformations  $f_i: M_i^3 \to O$, $i=1,2$, such that the pullback of the 
generator $z$ in \eqref{x3} is nonzero $f_i^* z=0$.
With $\pi_i : M_1^3 \times M_2^3 \to M_i^3$, $i=1,2$, the projection to the 
two factors, we consider the composite maps 
$g_i= f_i \circ \pi_i$. Then we use these maps to pull back $z$ as
$\alpha_i= g_i^* z$, $i=1,2$ and also define a third map  as the product
$g_3=g_1g_2$ with $\alpha_3=g_3^* z$. With this, $\alpha_3=\alpha_1 + \alpha_2$,
so that $q(\alpha_3)= q(\alpha_1) + q(\alpha_2) + 1$. This relation implies that,
upon using the change of framing formula \cite{JR}, one of the maps $g_i$, $i=1,2,3$ 
changes the value of the Kervaire invariant.
This demonstrates that the action in this example depends on the choice of framing.


%
%
%
%
%
%
%
%

%
%

\subsection{The connection to the Hopf invariant at the rational level}
\label{Rat}

We identify the topological part of the action of the M5-brane in seven dimensions 
with the Hopf invariant. We start at the rational level, making connection to \cite{Int},
 and then at the integral level in Section \ref{Int}
and also the $\Z_2$ level in Section \ref{Z2}. As far as we know, the latter two are new.
We consider the  M5-brane effective action lifted to seven dimensions,
alternatively viewed as a reduction of the M-theory action on a 4-manifold,
as in \cite{FSS1}. 
We will  
concentrate first on the case of the 7-sphere and then on more general 
7-manifolds. 
The relevant term is the topological coupling $\int C_3 \wedge F_4$,
which can be written as
$
\int *J\wedge C_3$,
i.e. as an interaction of the potential $C_{3}$ with the (conserved)
 topological current
$
J_{3} =  *_{7} F_4$ \cite{Tze}.

%

\medskip
For an arbitrary smooth map 
$f : S^7 \to S^4$, we describe
 the Hopf invariant $H(f)_\R \in \R$ 
in our context
as follows. 
This will be a straightforward generalization 
of the case of the second Hopf fibration presented in \cite{MMN}.
We start by choosing a 4-form $F_4 \in \Omega^4(S^4)$ on 
$S^4$  normalized as
$
\int_{S^4} F_4 =1
$.
This is satisfied, notably, for the Freund-Rubin ansatz \cite{FR}, where 
$F_4$ is taken to be constantly proportional to the volume form on $S^4$.
Then, since $d (f^* F_4) = f^* (dF_4)=0$ by the Bianchi identity, 
we have that $f^*F_4$ is a closed form on $S^7$. 
From $H^4(S^7;\R)=0$, by the de Rham theorem there exists 
a 3-form $C_3 \in \Omega^3(S^7)$ 
such that $f^*F_4 = d C_3$. Then the Hopf invariant of $f$ is defined as
\(
H(f)_\R=\int_{S^7} C_3 \wedge d C_3\;.
\)
The value of $H(f)_\R$ is determined 
independently of the choices of 
$F_4$ and $C_3$, and thus depends only on $f$. 
%
In fact, it can be shown that the value of $H(f)_\R$ depends only on the homotopy 
class of $f$, i.e., if two smooth maps $f_0, f_1: S^7 \to S^4$ are
homotopic, then $H(f_0)_\R=H(f_1)_\R$ (see \cite{MMN} for an illustration
for the similar second Hopf fibration).
 We highlight that the above statement  is  
 the gauge invariance of the 7-dimensional Chern-Simons term for $C_3$.

\medskip
If we replace $S^7$ with an arbitrary closed compact orientable 7-manifold 
$M^7$, we may still obtain an invariant of $f : M^7 \to S^4$ and we still have
\(
H(f)_\R = \int_{M^7} C_3 \wedge f^* \omega_4 =
\int_{M^7} C_3 \wedge F_4 =
\int_{M^7} C_3 \wedge d C_3\;,
\)
where $\omega_4$ is the volume 4-form on $S^4$ and $C_3$ satisfies
$F_4=f^*\omega_4=dC_3$. 
Then a straightforward generalization of the discussion for $S^7$ gives the 
following statement:
Let $M^7$ be a closed Riemannian extended worldvolume
 and $\omega_4 \in \Omega^4(S^4)$ 
be the area form on $S^4$. Then $H(f)_\R$ provides a homotopy invariant for 
a map $f : M^7 \to S^4$ if the 4-form $f^* \omega_4$ is exact. This may be viewed as a 
(generalized) Freund-Rubin ansatz. 


\paragraph{The Hopf invariant as a linking number.} 
The Hopf invariant is geometrically given by the {\it linking number} of the pre-images
$\ell_1=f^{-1}(r_1)$ and $\ell_2=f^{-1}(r_2)$ of two distinct regular values $r_1$ and $r_2$
of the map $f$. Choose open neighborhoods $W_{\ell_1}$ and $W_{\ell_2}$ of 
$\ell_1$ and $\ell_2$ 
and choose representatives $F_4^{\ell_1}$ and $F_4^{\ell_2}$ 
of the compact Poincar\'e duals of 
$\ell_1$ and $\ell_2$ in the cohomology groups with compact supports
 $H_c^4(W_{\ell_1})$ and $H_c^4(W_{\ell_2})$.
Now, since $H_{\rm dR}^4(S^7)=0$, the extensions 
of 
 $F_4^{\ell_1}$ and $C_3^{\ell_2}$ by zero to all of $S^7$ are exact, i.e.
 there are 3-forms $C_3^{\ell_1}$ and $C_3^{\ell_2}$ on $S^7$ 
 such that
 $dC_3^{\ell_1}= F_4^{\ell_1}$ and 
 $dC_3^{\ell_2}= F_4^{\ell_2}$.
The differential form definition of the linking number is 
\(
{\rm Linking~ number}=L(\ell_1, \ell_2)=\int_{S^7} C_3^{\ell_1} \wedge F_4^{\ell_2}\;.
\)
This is well-defined, as is shown e.g. in \cite{BT} 
for the case of $S^3$, and the proof for our case of $S^7$ is similar and, likewise, known.
%
%
%

\medskip
We now consider the case when our extended worldvolume is
a product of a 3-manifold 
$M^3$ and a 4-manifold $N^4$, relying on 
\cite{White} \cite{Tze}.  
Consider two continuous maps $f(M^3)$ and $g(N^4)$ from the two smooth,
oriented, non-intersecting manifolds $M^3$ and $N^4$ 
into Euclidean space $\R^{8}$. Let $S^7$ be the unit $7$-sphere 
centered at the origin of $\R^{8}$ and $e^* \Omega_{7}$ be the pullback 
of the volume form of $S^{7}$ under the map 
$e : M^3 \times N^4 \to S^7$, where we associate to each pair 
of points $({\bf m}, {\bf n}) \in M^3 \times N^4$ 
the unit vector $e$ in $\R^{8}$
given by $e({\bf m}, {\bf n})={(g({\bf n}) - f({\bf m}))}/
{|g({\bf n}) - f({\bf m})|}$.
The degree of this map is the generalized Gauss linking number of $M^3$ 
and $N^4$
\(
L\left( f(M^3), g(N^4)\right)
:= L(M^3,N^4) = \frac{1}{\Omega_{7}} 
\int_{M^3 \times N^4} e^* \Omega_{7}\;,
\)
which is a special case of the Gauss integral.
Note that $L(M^m,N^n)$ obeys the graded-commutativity law
$
L(M^m,N^n)=(-1)^{(m-1)(n-1)} L(N^n, M^m)$,
so that it is zero for even-dimensional manifolds $M^m$ and $N^n$,
and in our case this is symmetric so that it does not matter in 
which order we take $M^3$ and $N^4$.
\footnote{This is in contrast to the $f$-invariant that we consider in Section \ref{Sec f}.}
In the special case when 
we take $M^3=S^3$ and $N^4=S^4$ we get a linking number on
the product $S^3 \times S^4 \subset \R^8$, which can be viewed as
a result of `untwisting' the total space of the 7-sphere.  

\paragraph{Differential refinement of the Hopf invariant.}
The differential form that appears in the definition of the 
Hopf invariant has, in the context of M-theory, a refinement to 
a class in differential cohomology. Then the action, which is 
of Chern-Simons type, can be written as the refinement of the 
Hopf invariant 
\(
\hat{H}(f)=\int_{M^7} \hat{C}_3 \cup \hat{G}_4\;,
\)
which is exactly the type of expressions studied in \cite{FSS1}
\cite{FSS2} \cite{cup}. Therefore, the M5-brane action in seven dimension 
can be seen as a differential (and even  a stacky) refinement of the Hopf invariant.

%
%



\subsection{Hopf invariant 2 over $\Z$ and refinement via the dual of the C-field}
\label{Int}



We now consider the integral version of the Hopf invariant and relate it to 
the M-branes as well as to the C-field and its dual. 
It is also useful to extend to eight dimensions, as in \cite{Flux} 
\cite{HS}. 
 The Hopf invariant $H(f) \in \Z$ of a map 
$f : S^{7} \to S^{4}$  is determined by the cup product
structure of the mapping cone 
$
X=S^{4} \cup_f \D^{8}
$ 
with 
\(
a \cup a= H(f) b \in H^{8} (X; \Z)\;,    
\)
for generators $a \in H^{4}(X; \Z)=\Z$ and
$b \in H^{8}(X; \Z)=\Z$. 
Therefore, the Hopf invariant defines a morphism of groups
$
H : \pi_{7}(S^{4}) \to  \Z
$
given by 
$f \mapsto H(f)$\;. 
This can be characterized as follows.
Since $S^3$ is 2-connected, then the EHP exact sequence of homotopy 
\(
\xymatrix{
\cdots \ar[r]
&
\pi_6(S^3) 
\ar[r]^-\Sigma
&
\pi_7 (\Sigma S^3) 
\ar[r]^-{H}
&
\pi_6(S^3 \vee S^3) 
\ar[r]^-{P}
&
\pi_5(S^3)
\ar[r]
&
\cdots
}
\)
gives that the suspension $\Sigma$ and the Hopf invariant map $H$ fit into the
exact sequence
\(
\xymatrix{
\pi_{6}(S^3) 
\ar[r]^-\Sigma
&
\pi_{7}(S^{4})
\ar[r]^-{H}
 &
\Z
\ar[r]^-{P}
 &
 \pi_{5}(S^3)
 }\;.
\)
Note that the standard Hopf map
 $\eta : S^7 \to S^4$ 
 has Hopf invariant one: $H(\eta)=1$.  


\paragraph{Refinement of the Hopf invariant via the dual of the C-field}
We consider the dual $C_6$ of the C-field $C_3$  associated to the M5-brane
in a way that is `dual' to the way the C-field is associated to the M2-brane. 
The two fields $C_3$ and  $C_6$ are related as follows. 
The first is a 3-connection for a curvature $F_4$. At the level of 
differential forms, the Hodge dual $*_{11}F_4$ is a field $F_7$, which 
admits $C_6$ as a trivialization (or a 6-connection in the higher geometric language
of \cite{SSS1} \cite{SSS3}). 
The M2-brane has a coupling 
$\int_{M^3} C_3$, while the M5-brane has a coupling $\int_{M^6}C_6$.
The dual field is given in full generality by the right hand side of the equation of motion 
$d*F_4=\tfrac{1}{2} F_4 \wedge F_4 + I_8(g)$, where $*$ is the 
Hodge duality operator with respect to the metric $g_Y$ on the eleven-dimensional 
target space $Y^{11}$, and $I_8(g)$ is polynomial in the 
Pontrjagin forms of $Y^{11}$ given by $I_8=\tfrac{1}{48}(p_2(g) - (\tfrac{1}{2}p_1(g)^2))$. 
Lifted to cohomology, the corresponding degree eight class is \cite{DFM} 
\(
\Theta_Y= \tfrac{1}{2}a \cup a - \tfrac{1}{2} a \cup \lambda + 30 \hat{A}_8(Y^{11})\;,
\label{Theta}
\)
where $a$ is the class of an $E_8$ bundle on $Y^{11}$ and $\lambda$ is the first Spin 
characteristic class $\tfrac{1}{2}p_1$, related via the quantization 
condition $[F_4] + \frac{1}{2}\lambda=a \in H^4(Y^{11}; \Z)$ \cite{Flux}.

\vspace{3mm}
Now we observe that 
when the Pontrjagin classes vanish, or more precisely when we have 
String and Fivebrane structures (cf. \cite{SSS2}), 
then we get Hopf invariant 2
\(
a \cup a = 2 \Theta_Y\;.
\label{H2}
\)
Next, if we have the vanishing of the $\widehat{A}$-genus
 then 
we get 
\(
a \cup a -  a \cup \lambda= 2\Theta_Y\;.
\)
The right hand side is a quadratic refinement of the bilinear form on 
the cohomology group $H^4$ given by $\lambda$. Therefore, 
this suggests that 
we have a quadratic refinement of the Hopf invariant.
In the general case, we have the nonlinear term proportional to $\lambda^2$,
and so we view \eqref{Theta} as both a quadratic refinement 
and a nonlinear modification.

\paragraph{First example of the dual of the C-field and Hopf invariant 2.}
Let $f : S^{7} \to S^4$ be any continuous map. 
Consider the 8-dimensional cell complex 
$
X := \D^{8} \bigcup_f S^4
$
obtained by identifying a point on the boundary $x \in \partial \D^{8}=S^{7}$
with $f(x) \in S^4$. Thus $X$ is a CW complex with one cell each in 
dimensions $0, 4$, and 8. Then the cohomology groups of $X$ are 
$
H^i(X;\Z) \cong \Z$ if ${i=0, 4, 8}$ and are $0$ otherwise. 
Denote by $\sigma_i$ the generator of $H^i(X;\Z)$ determined by the cell
(endowed with an orientation) in dimension $i$ for 
$i=4, 8$. In terms of generators, the { Hopf invariant} of $f$ is the integer $H(f)$ such that
$
\sigma_4^2 =H(f) \sigma_{8}$.
If $H(f)=h \in \Z$ then the cohomology ring of $X$  is isomorphic to 
$
\Z[\sigma_4, \sigma_{8}]/
\langle \sigma_4^2=h \sigma_{8}, \sigma_4^3, \sigma_{8}^2
\rangle$.
In particular, if $h=1$, then $H^*(X;\Z)= \Z[\sigma_4]/\langle \sigma_4^3 \rangle$,
and this is the quaternionic projective  plane
$
{\mathbb H} P^2$. 
In our case, when we have Hopf invariant 2, the mapping cone is not 
a projective plane. 
It is obvious that we can allow $h$ to take any value we like; 
in particular, we can have the value $h=2$ appropriate for the dual of the
C-field. 

\paragraph{Second example with Hopf invariant 2.}
We consider the Hopf fibration and make use of two distinct copies of 
$S^4$. Take $S^4 \times S^4$ as the cell complex formed by attaching 
an 8-cell to the wedge of two spheres $S^4 \vee S^4$, using 
the attaching map $S^7 \to S^4 \vee S^4$. This can be described using 
the Whitehead product, by forming the composition of $g$ with 
the folding map $F: S^4 \vee S^4 \to S^4$. Starting with two base-point preserving maps
$f: S^4 \to X$ and $g: S^4 \to X$, let 
$[f, g]: S^7 \to X$ be the composition 
$
\xymatrix{
S^7 
\ar[r]
& 
S^4 \vee S^4 \ar[r]^{~~f \vee g}
&
X
}
$. 
This gives a well-defined product $\pi_4(X) \times \pi_4(X) \to \pi_7(X)$,
which generalizes the commutator product for the (nonabelian)
fundamental group. 
Then the map $S^7 \to S^4_\alpha \vee S^4_\beta$
is the Whitehead product $[\iota_\alpha, \iota_\beta]$ of the two inclusions 
of $S^4$ into $S^4_\alpha \vee S^4_\beta$. 
Now let $X^8=e^8 \bigcup_{[\iota, \iota]}S^4$ be the space obtained from $S^4$ by 
attaching an 8-cell $e^8$ via the map representing 
$[\iota, \iota]$. Let $u \in H^4(X^8; \Z)$ and $v \in H^8(X^8; \Z)$ be 
the cohomology generators in degree four and eight, respectively. 
Then, if we take $\iota$ to be the class of the identity map, then 
$u^2=2v$, so that the Whitehead product $[\iota, \iota]$ has Hopf invariant $\pm 2$. 

\paragraph{Third example with Hopf invariant 2.}
Let $U^4_+$ be the northern hemisphere of $S^4$ and let
$g: (U^4_+, S^3) \to (S^4, (1,0,0,0))$ be any map whose 
restriction to the open upper-hemisphere $U^4_+ -S^3$ is a homeomorphism.
Now view $S^7$ as the boundary of $\mathbb{D}^4 \times \mathbb{D}^4$. 
On the other hand, the simplicial boundary of $\mathbb{D}^4 \times \mathbb{D}^4$
is also $\partial \mathbb{D}^4 \times \mathbb{D}^4 + \mathbb{D}^4 \times \partial \mathbb{D}^4$.
Then define the map 
$
f: \partial \mathbb{D}^4 \times \mathbb{D}^4 + \mathbb{D}^4 \times \partial \mathbb{D}^4
\to
S^7$
by 
$
(y, (x_1, x_2, x_3, x_4)) \in \partial \mathbb{D}^4 \times \mathbb{D}^4 \longmapsto  
g\left(x_1, x_2, x_3, x_4, \sqrt{1-(x_1^2 + x_2^2 + x_3^2 + x_4^2)}\right)
$ on the first factor and 
$
((y_1, y_2, y_3, y_4), x) \in  \mathbb{D}^4 \times \partial \mathbb{D}^4 \longmapsto
g\left(y_1, y_2, y_3, y_4, \sqrt{1-(y_1^2 + y_2^2 + y_3^2 + y_4^2)}\right)
$
on the second factor.
Then this map $f$ has degree 2 and hence Hopf invariant 2. 

\vspace{3mm}
\noindent {\bf Remarks.}
The Hopf invariant  is $H(f)=-2$ if we choose the orientation of $S^7$ determined by the
generator $-a$ instead of $a$. Moreover, 
starting with a map with Hopf invariant one, we can get a map of 
Hopf invariant $k$ by composing with a map on the seven sphere 
of degree $k$. 
Let $h : S^7 \to S^7$ be a map of degree $k$, i.e. $H^*(a)=ka$,
and $f : S^7 \to S^4$ a continuous map. Then
$
H(fh)=k~ H(f)$. Taking $k=2$ we get a map with $H(f)=2$. 
Note that, in contrast, this cannot be generated by composing with 
a map of degree 2 on the 4-sphere. 
Indeed, let $f: S^7 \to S^7$ be a continuous map, and 
$h : S^4 \to S^4$ a map of degree $k$, i.e. $h^*(u)= ku$. Then
$
H(hf)= k^2 H(f)
$ and, of course, $k^2$ cannot be 2.


\subsection{M-branes via cohomotopy}
\label{Sec coh}

The topological study of the C-field amounts to an 
interpretation of a degree three cohomology class together
with some extra structure. One could consider this as an 
$E_8$ bundle or as a $K(\Z, 3)$ bundle. 
The latter can be viewed as an approximation of the former
in the range of dimension of M-theory. 
We will propose another interpretation, whereby we use the 
loop space $\Omega S^4$ of the 4-sphere. Note that 
$BE_8 \simeq K(\Z, 4)$ and that $\pi_4(K(\Z, 4))=\pi_4(S^4)$
with obvious injection $\pi_i(K(\Z, 4)) \to \pi_i(S^4)$. 
Therefore, we can account for the degree four class 
corresponding to the class of $F_4$ using the 4-sphere 
as a model. However, we will see that the connection 
is much more precise.
We have a degree four field $F_4$ and a degree seven field 
$F_7:=*F_4$, 
satisfying the Bianchi identity and the equation of motion of the C-field 
(without correction terms)
\(
dF_4=0\;, 
\qquad \qquad  \qquad
d*F_4=\tfrac{1}{2}F_4 \wedge F_4\;.
\label{Eq for F}
\)

\paragraph{Rational homotopy of $S^4$.}
Consider the 4-sphere $S^4$ with its volume form $\omega_4$.
The de Rham cohomology $H^*_{\rm dR}(S^4)$ is not
an exterior algebra, and one has to impose the condition 
$\omega_4 \wedge \omega_4=0$.  
Correspondingly, let $x_4$ be the generator of de Rham cohomology
in degree four. In addition to $dx_4=0$, this generator has to be 
such that $x_4^2$ is zero in the cohomology of the minimal model.
This means that there is a generator $y_7$ in degree seven
such that a differential $d$ can be defined with
$x_4^2=dy_7$. This then gives that $x_4$ and $y_7$ are the 
only generators because of the above imposed relation $x_4^2=0$ and the 
automatic relation $y_7^2=0$. Sullivan's minimal model (see \cite{FHT}) for 
$S^4$ is then given by the map 
$\tau: \left( \bigwedge (x_4, y_7), d\right) \longrightarrow \mathcal{A}(S^4)$,
where $dy_7=x_4^2$ and $\tau (x_4)$ represents the unique 
cohomology generator of degree eight. The map from the minimal model
to the de Rham complex of $S^4$ is given by sending $x_4$ to the volume form
$\omega_4$ of $S^4$ and sending $y_7$ to zero. 
The point here is that these expressions correspond to equations 
\eqref{Eq for F}.

\medskip
Note that $\tau$ is a quasi-isomorphism of free commutative 
differential graded algebras so that the rational homotopy
$\pi_* (S^4) \otimes \Q$ is concentrated in degrees 4 and 7; furthermore, 
the rank in each of these degrees is one. In fact, one can consider the Whitehead
product $\pi_4 (S^4) \otimes \pi_4 (S^4) \to \pi_7 (S^4)$, 
which we used in the second example in Section \ref{Int}, and which is
given by 
$f, g \longmapsto [f, g]$, where $f, g: S^7 \to S^4$ are the 
corresponding maps. 
Since $S^4$ is simply-connected then the vector space
$\pi_*^\Q(S^4):=\pi_i^\Q(S^4)$ 
turns into a graded Lie algebra with the Whitehead product as the Lie bracket
of degree $-1$.

\paragraph{The rational 4-sphere.}
A simply connected space $X$ is rational if $\pi_*(X)$, $\widetilde{H}_*(X; \Z)$, or $\widetilde{H}_*(\Omega X; \Z)$ is 
a $\Q$-vector space, with the three conditions being equivalent, where $\Omega X$ is the space of 
based loops on $X$. 
Then the rational 4-sphere is defined to be (see \cite{Hess}) 
$S^4_\Q:= \left( \bigvee_{k\geq 1} S_k^4 \right) \bigcup \left( \coprod_{k \geq 2} \mathbb{D}_k^5 \right)$,
where $\mathbb{D}_k^5$ is attached to $S^4_k \vee S^4_{k+1}$ via a 
representative $S^4 \to S^4_k \vee S^4_{k+1}$ of $\iota_{4, k} - (k+1)\iota_{4, k+1}$, where 
$\iota_{4,k}$ denotes the homotopy class of the inclusion of $S^4$ as 
the $k$-th summand $S^4_k$ of $\bigvee_{k \geq 1} S^4_k$. Then 
the (reduced) homology of this space is $\widetilde{H}_*(S^4_\Q; \Q)=\Q$
for $k=4$ and zero otherwise.

\paragraph{Cohomotopy.}
There is the Borsuk cohomotopy functor $\pi^M= [-, M]: {\rm \bf Top} \to {\rm \bf Set}^{\rm op}$
from the category of topological spaces to the (opposite) category of sets, given by 
$\pi^M(X)=[X, M]$, where
$M$ is an arbitrary topological space. This is a homotopy invariant. 
When $M=S^n$, an $n$-sphere, then this is the usual cohomotopy
functor $\pi^n(X)=[X, S^n]$. 
When $X$ is of dimension less or equal to $2n-1$ then the set 
$\pi^n (X)$ has a canonical abelian group structure arising from taking suspension
$\Sigma: [X, S^n] =[\Sigma X, S^{n+1}]$
and using suspension coordinates; see \cite{Hu}. 

\medskip
We propose that the C-field and its dual on $Y^{11}$ are captured by the 4-sphere, 
via homotopy classes of maps $[Y^{11}, S^4]$. This cohomotopy set $\pi^4(Y^{11})$
is in canonical bijection with the set of cobordism classes of codimension-4
framed submanifolds $N^7$ of $Y^{11}$. The set $\pi^4(Y^{11})$ is in general 
not a group, but we will take $Y^{11}$ to have a topologically nontrivial factor 
with relatively low dimension. In order to be in the stable range, we will take
$Y^{11}=N^7 \times \R^4$, so that the homotopy classes of maps become
$[N^7, S^4]=\pi^4(N^7)$. When we take $N^7=S^7$ then, by duality, we 
have $\pi^4(S^7)=\pi_7(S^4)$. This reduces to the discussion above on the Hopf invariant.
In terms of
the sphere spectrum ${\cal S}$ we have the stable cohomotopy
\(
\pi_s^{-*}(X)=\pi_* [X, {\cal S}]\;.
\)

\paragraph{Relation to the Pontrjagin-Thom construction.}
Let ${\cal F}_7(Y^{11})$ denote the set of closed $7$-dimensional 
submanifolds of $Y^{11}$ with a framing on their normal bundle, up to 
normally framed bordism in $Y^{11} \times [0,1]$.
We have the following diagram 
\(
\xymatrix{
\pi^4(Y^{11})=[Y^{11}, S^4] 
\ar[rr]
\ar[d]^{h^4}
&&
{\cal F}_7(Y^{11})
\ar[d]^{h_7}
\\
H^4(Y^{11}) 
\ar[rr]
&&
H_7(Y^{11})\;,
}
\)
where the map $h^4$ pulls back the cohomology class of $S^4$. 
The forgetful map $h_7$ uses the normal framing to orient the 
submanifold $N^7$ and then push forward its 
homology fundamental class.

\paragraph{Examples.} We can take $Y^{11}$ to be of the form 
${\rm AdS}_4 \times S^7$ or ${\rm AdS}_7 \times S^4$. If we take the anti de Sitter factors to be
Euclideanized, ${\rm AdS}_i^{\rm Euc}$, then the 
cohomotopy $[Y^{11}, S^4]$ becomes homotopy equivalent to
 $[S^7, S^4]=\pi_7(S^4)=\Z$ and 
$[S^4, S^4]=\pi_4(S^4)=\Z$. 
One can also allow for more general situations leading to fractions of 
supersymmetry, namely to allow for more general Einstein manifolds 
$M^4$, or even orbifolds $M^4/\Gamma$, where $\Gamma$ is a 
finite subgroup of SO(4), most notably a cyclic group $\Z_k$ 
of order $k$. Then the Hopf isomorphism for cohomotopy (dual to the Hurewicz isomorphism
for homotopy) gives
\(
\pi^4(Y^{11})=[{\rm AdS}_7^{\rm Euc}\times M^4, S^4] \cong H^4(M^4)\;.
\)
For example, with $M^4=S^4/\Z_2=\R P^4$, this captures the 
2-torsion part of the C-field, as in e.g. \cite{Flux} \cite{Ho}.

\section{M-branes as corners and topological invariants}
\label{Sec corner}

We now consider corner structures on the M2-brane and the M5-brane,
as in \cite{corner} \cite{boundary}.
This uncovers some structures and connections to 
invariants, such as the $\Z_2$ Hopf 
invariant  and the $f$-invariant. 
In addition, we will consider the connection to a geometric/topological invariant, 
namely the $\nu$-invariant for 7-manifolds with a  $G_2$-structure.

\subsection{M-branes and corners}
\label{M co}

We start with the M2-brane and then consider the M5-brane. In each 
case we characterize the corner structure as it relates to the topological
action.

\paragraph{The M2-brane and corners.}
We have considered the 3-dimensional worldvolume $M^3$ as the boundary of a 4-dimensional 
manifold $W^4$ in order to consider anomalies via the field strength $F_4$, as in 
\cite{Flux}. 
On the other hand, we have also considered 
boundaries of M2-branes -- effectively the self-dual strings-- ending on the 
M5-brane worldvolume. As in the approach of \cite{corner} \cite{boundary} \cite{F},
putting the two together leads to viewing the 
self-dual string as a corner of codimension-2. 
The structure of the manifolds involved and corresponding topological actions take the form
\(
\xymatrix{
W^4: 
\ar[d]^{\partial_1}
& \int_{W^4} F_4\;, && {\text{Manifold~with~corners~of~codimension-$2$.}} \\
M^3: 
\ar[d]^{\partial_2}
& \int_{M^3} C_3\;, && {\rm Boundary.}\\
\Sigma_2: & \int_{\Sigma_2} B_2 && {\text{Corner~of~codimension-$2$.}}
}
\)

\paragraph{The M5-brane and corners.}
We have discussed in \cite{corner} \cite{boundary} 
how the M5-brane worldvolume can be viewed as a corner, essentially 
due to the  presence of tubular neighborhood. 
The main source of spaces with corners of codimension-2 is the 
product of two manifolds with boundary. We will focus on the representative case
of the product of two closed disks.
The fivebrane worldvolume, taken as $S^3 \times S^3$, can be viewed as 
a corner in two different ways, schematically 
\(
\xymatrix{
&&
\D^4 \times \D^4 
\ar[dl]_{\partial_1}
\ar[dr]^{\partial_2}
&&
\\
&
S^3 \times \D^4 
\ar[dl]_{\partial_2}
\ar[dr]^{S^1}
&&
\D^4 \times S^3
\ar[dl]_{S^1}
\ar[dr]^{\partial_1}
&
\\
S^3 \times S^3 
&
&
S^2 \times \D^4
&
&
S^3 \times S^3\;.
}
\)
The corresponding topological action decomposes accordingly. 
Starting with the quadratic action in eight dimensions and 
keeping track of two different components we have a reduction 
pattern schematically of the form
\footnote{At the face of it, it seems like we have a major problem:  we need Stokes theorem for manifolds 
with corners, in the sense of having double boundary operators on the manifolds 
and double exterior derivatives  the corresponding forms,  which we do  not have.
However, we approach the corners from the topology point of view via cobordism,
which avoids the need for boundary operator calculus, which may or may not exist;
see \cite{corner}\cite{cup}\cite{F}. The main point is that higher connections 
can be themselves be viewed as curvatures. 
} 
\(
\xymatrix{
&&
F_4 \cup F_4' 
\ar[dl]_{\partial_1}
\ar[dr]^{\partial_2}
&&
\\
&
C_3 \cup F_4' 
\ar[dl]_{\partial_2}
\ar[dr]^{S^1}
&&
F_4 \cup C_3'
\ar[dl]_{S^1}
\ar[dr]^{\partial_1}
&
\\
C_3 \cup C_3' 
&
&
B_2 \cup F_4' ~~{\rm or}~~ H_3 \cup C_3'
&
&
C_3 \cup C_3'\;,
}
\)
where we take $B_2=\pi_* (C_3)$ or $H_3=\pi_*(F_4)$ obtained by integration 
over the circle fiber (as in the 11-dimensional case, see \cite{MS}). 
The entries in the middle row correspond to Chern-Simons extensions and are examples of 
(higher) cup-product Chern-Simons theories \cite{cup} \cite{up}. 
The extension of such theories to the bottom line, corresponding to the corners, is currently 
being developed.

\subsection{The $\Z_2$ Hopf invariant}
\label{Z2}

In Section \ref{Rat}  and Section \ref{Int}
 we considered the relation to the 
rational and integral Hopf invariants, respectively. Here
we start with the relation of the mod 2 Hopf invariant to both the 
M2-brane and the 
M5-brane. 

\paragraph{The $\Z_2$ Hopf invariant and the M2-brane.}
The mod 2 Hopf invariant   $H_2(g)\in \Z_2$ of a map 
$g : S^7 \to S^4$ defines an isomorphism 
$
H_2 : \pi_{5}(S^4) \to  \Z_2$
given by
$g \mapsto  H_2(g)$. 
 This is determined by the Steenrod square on the
mod 2 cohomology of the mapping cone
$
X^8=S^4 \bigcup_g \D^{8}
$ 
with 
\(
Sq^4=H_2(g): H^4(X^8;\Z_2)\cong \Z_2 
\longrightarrow H^{8}(X^8;\Z_2)\cong \Z_2\;.
\)
In this case, 
the mod 2 Hopf invariant $H_2(g) \in \Z_2$ is the mod 2 reduction of
the integral Hopf invariant $H(g) \in \Z$. 
We will make use of the nice description for this invariant given in \cite{Mi}.

\vspace{3mm}
Consider a vector bundle $E\to M$ associated to 
the $O(n)$-universal bundle $EO(n) \to BO(n)$ over the classifying space of the
orthogonal group $O(n)$. 
The corresponding Thom spectrum is  
\(
MO(n):=(BO^E)_n={\cal T}(\R^n \oplus E) \cong S^n \wedge {\cal T}(E) \cong \Sigma^n {\cal T}(E)\;.
\)
The inclusions $O(n) \hookrightarrow O(n+1)$ induce a directed system of such 
spectra, and the Thom spectrum 
 $MO$ is the direct limit $MO:=\lim_\to MO(n)$. 
  An element of the homotopy group
   $\pi_n(MO)$ is naturally identified with the set of 
cobordism classes of closed $n$-manifolds.
Now consider the quotient $\overline{MO}=MO/\cS$ by the 
sphere spectrum  $\cS$, which is the suspension spectrum of a point
and which detects whether or not the manifold is framed. 
Of interest to us is 
an element of the fourth homotopy group of this space 
$\pi_4(\overline{MO})$, which represents a class of triples
$(W^4, M^3, f)$ in which $W^4$ is a 4-manifold with boundary $M^3=\partial W^4$, and 
$f$ is a trivialization of the normal bundle $\nu_M$. 
As before, $M^3$ represents the worldvolume of the M2-brane, 
$W^4$ is its bounding 4-manifold, and the normal bundle ${\cal N}_N$ to $W^4$ 
can be identified with the 
normal bundle in the embedding in 11-dimensional spacetime $Y^{11}$, as here
we are already in the stable range.

\medskip
Such an $(O, {\rm fr})$-manifold represents zero if it is 
a boundary, i.e. if it embeds in a 5-dimensional 
manifold with corner $(N^5, W^4, W'^4, M^3, f)$. This means that $N^5$ is a 
5-manifold whose boundary is given by 
$W^4 \bigcup_{M^3}W'^4$, with $W^4$ and $W'^4$ are manifolds with 
boundary the 3-manifold $M^3$, $\partial W^4=M^3=\partial W'^4$,
and $f'$ is a trivialization of the normal bundle of $W'^4$ which restricts to the
given trivialization of the normal bundle of $M^3$. 
The map $\pi_4(\overline{MO})\to \pi_4(\cS)$ sends 
$(W^4, M^3, f)$ to its boundary 
$(M^3, f)$, with $f=f'|_{M^3}$
\(
\xymatrix{
W^4 
\ar@{..>}[rr]^{f'}
\ar@{..>}[drr]^{\hspace{-1cm}\alpha'}
&&
EO
\ar[d]
&
O
\ar[l]
\\
M^3
\ar@{_{(}->}[u]^\iota
\ar[urr]^{\hspace{11mm}f}
\ar[rr]^\alpha
&&
BO\;.
}
\)
We now consider classes  in $\overline{H}^*(BO; \Z_2)$ corresponding to the 
spectrum $\overline{MO}$. 
Let us denote by ${\cal N}={\cal N}_N$  the normal bundle of $W^4$. The trivialization $f$ of ${\cal N}|_{M^3}$ 
provides a factorization of $W^4 \to BO$ through $W^4/M^3$.
Hence, as explained more generally in \cite{Mi},  
any $c \in \overline{H}^k(BO; \Z_2)$
gives rise to a relative class $c({\cal N}, t) \in H^k(W^4, M^3; \Z_2)$. In particular, 
we consider the relative Stiefel-Whitney class
$w_4({\cal N}, f)\in H^4(W^4, M^3; \Z_2)$. The $\Z_2$ Hopf invariant is
captured by the Hurewicz map on $\pi_4(\overline{MO})$
\cite{Sto}. In our case, 
\(
{\rm Hopf}(M^3, f)=
\langle 
w_4({\cal N}_{W^4},f), [W^4, M^3]
\rangle\;.
\)
Note that the Stiefel-Whitney class $w_4$ is related to the first Spin characteristic 
class $\lambda=Q_1=\tfrac{1}{2}p_1$ via mod 2 reduction; this is used 
extensively in \cite{tcu} \cite{tcu2} \cite{top1} \cite{top2}.
Thus we have 
\bea
{\rm Hopf}(M^3, f)=0 &\Leftrightarrow & w_4({\cal N}_{N}, f)=0
\nonumber \\
&\Leftrightarrow & \lambda({\cal N}_{N}, f)\in 2\Z
\nonumber\\
&\Leftrightarrow & \tfrac{1}{2}\lambda ({\cal N}_{N}, f)\in \Z\;.
\eea
We see that this captures the divisibility of $\lambda$ by 2.
Therefore, we arrive at the conclusion that the quantization 
condition on the C-field, $[F_4]+ \tfrac{1}{2}\lambda\in \Z$, holds 
if the Hopf invariant vanishes, 
${\rm Hopf}(M^3, f)=0$.

\paragraph{The $\Z_2$ Hopf invariant and the M5-brane.}
We now consider the situation for the M5-brane; the argument here
 will essentially be a `shift up of degree four' of the argument 
 presented for the 
M2-brane above. That is, we will deal with manifolds of 
dimensions 6, 7, and 8, replacing those of dimension 
2, 3, and 4, respectively. However, unlike the case of the M2-brane
which could be taken as the boundary, the M5-brane will in fact be the 
corner, i.e. the bottom manifold in the above hierarchy of 
three manifolds of consecutive dimensions.  
We consider the M5-brane worldvolume as the boundary of 
a seven-manifold $N^7$, which in turn is the boundary of an
eight-manifold $W^8$, making the latter a manifold with corners 
of codimension-2.  
The Hopf invariant in this case is 
\(
{\rm Hopf}(N^7, f)= \langle w_8({\cal N}_N , t), [W^8, N^7]\rangle\;.
\)
On the other hand, the one-loop polynomial  that appears in the 
degree eight cohomology class 
\footnote{See expression \eqref{Theta} and the discussion around it.} 
$\Theta$
 can be written in terms of the second 
Spin characteristic class $Q_2$ as \cite{KSpin}
\(
I_8=\tfrac{1}{24}Q_2\;,
\label{Q2}
\)
where we have $w_8=Q_2$ mod 2. This means that if $I_8$ is integral,
which is the case for Spin ten-manifolds with $w_4=0$ (see \cite{Flux}), 
then certainly $Q_2$ is divisible by 2, and hence $w_8=0$. This then
gives zero Hopf invariant. In the general case, the requirement is that the 
second Spin characteristic class $Q_2$ is divisible by 2. This can be 
viewed as an analog of the requirement of the first Spin characteristic 
class $Q_1=\tfrac{1}{2}p_1$ to be divisible by 2, in order for the 
flux quantization condition on the C-field to hold precisely.

\medskip
Note that if ${\cal N}_N$ admits a Fivebrane structure, in the sense of 
 \cite{SSS2} \cite{SSS3}, then $I_8$ is zero in cohomology, and
we have 
\eqref{H2},
so that the dual of the C-field defines a Hopf invariant 2.
As we indicated earlier, 
in general  the dual of the C-field provides a refinement of the 
Hopf invariant.

\subsection{The M5-brane and the Kervaire invariant}
\label{Ker}

In Section \ref{M co} we motivated the corner structure for the M-branes
and described how the corresponding topological action decomposes in various 
dimensions. Here we continue the discussion by
seeking an explicit relation to the Kervaire invariant. 
To that end, we will consider the 6-dimensional worldvolume 
for the M5-brane, viewed as a corner/codimension-2 defect, 
while for the M2-brane the corner will be the 2-dimensional boundary 
of the corresponding worldvolume.

\medskip
Consider again the reduced Thom spectrum $\overline{MO}=MO/\cS$.
We are interested in considering classes in the 
wedge sum $\overline{MO} \vee \overline{MO}$; this is the quotient of the 
disjoint union of 
two copies of $\overline{MO}$ by the equivalence obtained by identification of points. 
In the case of the sphere, this operations gives a bouquet of spheres.
\footnote{There is an analogous discussion in the case of Eilenberg-MacLane spectrum
in \cite{DMW} (section 3.2), which can be recast in the above language.}
An element of the homotopy group $\pi_8(\overline{MO}\vee \overline{MO})$
is represented by a $({\rm O}, {\rm fr})^2$-manifold, i.e. a framed manifold with corners of 
codimension 2. This consists of the
data $(W^8, N_1^7, N_2^7, {\cal N}_1, {\cal N}_2, f_1, f_2)$,
where $W^8$ is an 8-manifold with boundary $\partial W^8=N^7=N_1^7 \bigcup_{M^6}
N_2^7$, with corner $\partial N_1^7=M^6=\partial N_2^7$. The normal 
bundle ${\cal N}_{W^8}$ comes with a splitting 
${\cal N}_{W^8}={\cal N}_1 \oplus {\cal N}_2$, and here $f_1$ is a trivialization of ${\cal N}_1|_{N_1^7}$
and $f_2$ is a trivialization of ${\cal N}_2|_{N_2^7}$. The normal bundle of the
corner $M^6$ thus acquires a trivialization $f$ as well. 
The map $\pi_8(\overline{MO}\vee \overline{MO}) \to \pi_7(\overline{MO})$
carries the above data to $(N_1^7, M^6, f)$. 

\paragraph{The M5-brane and the Kervaire invariant.}
Let $(W^8, N^7_1, N^7_2, {\cal N}_1, {\cal N}_2, f_1, f_2)$ be 
an $8$-dimensional $(O, {\rm fr})^2$-manifold representing the
extended M5-brane worldvolume, as in \cite{corner} \cite{boundary} \cite{F}. 
Then, by the Lannes-Miller theorem \cite{La} \cite{Mi},  
the Kervaire invariant is given by 
\(
{\rm Kervaire}(M^6, f)=\sum_{i=0}^3 
\langle
v_{4-i}({\cal N}_1, f_1) \cup v_i({\cal N}_2) \cup v_{4}({\cal N}_2, f_2), 
[W^8, N^7]
\rangle\;,
\)
where $v_j$ is the $j$-th Wu class. 
We take ${\cal N}_1$ and ${\cal N}_2$ to be (at least) oriented, 
i.e. $v_1({\cal N}_1, f_1)=0=v_1({\cal N}_2, f_2)$,
so that there are two contributions 
$v_4({\cal N}_1, f_1) \cup v_4({\cal N}_2, f_2)$ and 
$v_4({\cal N}_1, f_1) \cup v_2^2({\cal N}_2, f_2)$
in the above sum, that is 
\(
\left\langle
 w_2({\cal N}_1, f_1) w_2({\cal N}_2) (w_4 ({\cal N}_2, f_2)
 + w_2^2 ({\cal N}_2, f_2)),
 [W^8, N^7]
\right\rangle\;.
\)
If we, furthermore, 
assume the Spin condition
\footnote{The distinction between orientation and 
Spin structure on the M5-brane has an interpretation via
elliptic cohomology \cite{KS1}.}
then $v_4=w_4$ and so  
\(
{\rm Kervaire}(M^6, f)=
\langle
w_4({\cal N}_1, f_1) \cup w_4({\cal N}_2, f_2),
[W^8, N^7] 
\rangle\;.
\)
If $w_4$ of either ${\cal N}_1$ or ${\cal N}_2$ is zero then this Kervaire invariant 
vanishes. One implication of this is that the action (see \cite{Eff} \cite{HS})
\(
\int_{W^8} F_4 \cup F_4 + \lambda \cup F_4 \mod 2
\) 
becomes only quadratic without a refinement. 



%
%
%
%


\paragraph{The M2-brane with boundary and the Kervaire invariant.}
Now consider the boundary of the M2-brane as a corners, as in \cite{corner}. 
That is, we have a 2-manifold $X^2$, which is the boundary of the 
M2-brane worldvolume $M^3$. This in turn is the boundary of a 4-manifold, in the
above sense. 
The self-dual string $\Sigma=X^2 \subset M^6$ lies inside the M5-brane worldvolume.
Then, assuming the normal bundles to be oriented we have 
\(
{\rm Kervaire}(X^2, f)=
\langle
w_2({\cal N}_1, f_1) \cup w_2({\cal N}_2, f_2),
[W^4, M^3] 
\rangle\;.
\)
We see that a Spin structure on either factor, ${\cal N}_1$ or ${\cal N}_2$,
is equivalent to the vanishing  of the 
Kervaire invariant of the self-dual string. 
Note, furthermore, that in the sigma model description
spinors take values in the normal bundle, and hence the above 
statement is a natural one to have. 

\subsection{M-branes and the Maslov index}
\label{Mas}

We have seen that the M2-brane action can be described in terms of 
the signature of a bounding 4-manifold. In the case of the M5-brane we
have something similar: the effective action and partition function of the 
M5-brane can be described via the signature in eight dimensions, as
described in \cite{HS}, which builds on \cite{Eff}. The relevant term 
in the partition function is \cite{HS} \cite{BM}
\(
\exp \left[2\pi i k \tfrac{1}{8} \int_{W^8}{\check \lambda} \cup {\check \lambda} - L_8   \right]\;,
\label{PF}
\) 
where $L_8$ is the degree eight component of the Hirzebruch 
L-polynomial, and ${\check \lambda}$ is a differential integral 
lift of the degree four Wu class $v_4$.
When $W^8$ is a manifold with boundary, then the contribution to the 
boundary is given by the signature defect, which is essentially
(a differential refinement of) the eta-invariant. 
We study consequences to this formula of having a corner instead. 
Note that the $f$-invariant is the right answer for the replacement of the 
eta-invariant for manifolds with corners. However,
before explicitly considering the $f$-invariant, we will make connection 
to the Maslov index. 
\footnote{Comprehensive literature and references on the Maslov index can be found in \cite{Ran}.}
This can be viewed as the signature defect in the presence
of corners, which leads to identifying the contribution to the action due to the 
presence of corners. 


\paragraph{The Maslov correction to the M2-brane.}
Now consider the M2-brane worldvolume $M^3$, viewed as a boundary of 
a compact 4-manifold $W^4$. In fact, 
we take three copies $M_i^3$ , $i=0,1,2$ of $M^3$ and 
 take $W^4$ 
to be decomposed along a component $M_0^3$ of $M^3$ into 
two parts $W_1^4$ and $W_2^4$ such that 
\(
\partial W_1^4= (-M_0^3) \cup M_1^3\;,
\qquad
\partial W_2^4= (-M_0^3) \cup M_2^3\;,
\qquad
\partial M_0^3=\partial M_1^3= \partial M_2^3=X^2\;. 
\)
Then the signature of $W^4$ is given in terms of the signatures of 
$W_1^4$ and $W_2^4$ and a correction term $\mu$ that arises from 
the way the homology classes of the spaces  involved are related. 
That is,
\(
{\rm sign} (W^4) = {\rm sign}(W_1^4) + {\rm sign}(W_2^4) 
+ \mu (K_0, K_1, K_2)\;,
\)
where $K_i=\ker \left(H_1(X^2; \R) \to H_1(M_i^3;\R) \right)$
for $i=0, 1,2$; so we are considering nontrivial 
1-cycles in the self-dual string 
$X^2$ that become homologically 
trivial when lifted to $M^3$. 
The importance of this, from a physical point of view, is that the 
dynamics of the BPS states in Seiberg-Witten theory \cite{SW} for which 
$X^2$ is the defining curve, are governed by such cycles. 
The Maslov index $\mu (K_0, K_1, K_2)$ is defined as 
follows.
%
%
We will restrict $H_1(X^2)$ to the domain of 1-cycles
satisfying $c_0 + c_1 + c_2$, with $c_i\in K_i$, $i=0,1,2$,
that is, giving a zero total cycle. This domain is the intersection 
$K_0 \cap (K_1 + K_2)$. Now we would like to quotient this 
space by 1-cycles that are made of pairs of 1-cycles that 
sum to the zero 1-cycle, i.e. by elements of $K_0 \cap K_1$ and 
of $K_1 \cap K_2$ with $c_0 + c_1=0$ and $c_0 + c_2=0$, respectively, 
with obvious symmetries.
Then we form the vector space 
\(
V=\frac{K_0 \cap (K_1 + K_2)}{(K_0 \cap K_1) + (K_0 \cap K_2)}
\label{W}
\)
with elements denoted by $[c_0]$, representing those 1-cycles
that satisfy $c_0 + c_1 + c_2=0$. Corresponding to the skew-symmetric
bilinear form $\omega$ on the first homology group
$H_1(X^2)$ is a symmetric bilinear form on $V$
defined by 
$
\rho ([c_0], [c_0'])= \omega (c_0, c_1')$.
The Maslov index is then the signature of this form
\(
\mu (K_0, K_1, K_2):= {\rm sign} (\rho)\;.
\label{mu}
\)
This is the contribution to the topological action of the M2-brane 
arising from the Maslov index.



\paragraph{The Maslov correction to the M5-brane.}
We now consider the M5-brane in the setting that we have
had: as a corner of codimension-2 of an 
8-dimensional manifold $W^8$. A similar study from an analytic point of
view is taken in \cite{corner}.
The signature of an oriented 8-manifold with boundary 
$(W^8, N^7)$ is the signature of the symmetric intersection form
on $H^4(W^8; \R)$, i.e. ${\rm sign} (W^8)={\rm sign} (H^4(W^8; \R))\in \Z$.
We now decompose $W^8$ in into three pieces 
$W^8_i$, $i=1,2,3$, such that $W^8=W_1^8 \cup W_2^8 \cup W_3^8$
is the union of the three codimension-0 manifolds with boundary meeting 
transversally with 
$
W_1^8 \cap W_2^8 \cap N^7 = 
W_2^7 \cap W_3^8 \cap N^7=
W_3^7 \cap W_1^8 \cap N^7=\emptyset
$.
The corner $M^6:=W_1^8 \cap W_2^8 \cap W_3^8$
is a codimension-2 submanifold and hence is identified as the
M5-brane worldvolume

\medskip
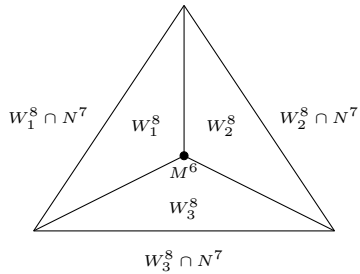
\begin{center}
\begin{tikzpicture}
\draw (-1,0) -- (1,3);
\draw (1,3) -- (3,0);
\draw (-1,0) -- (3,0);
\draw (1,3) -- (1,1);
\draw (1,1) -- (3,0);
\draw (1,1) -- (-1,0);
\node at (1.5,1.4) {\tiny $W_2^8$};
\node at (0.5,1.4) {\tiny $W_1^8$};
\node at (1,.3) {\tiny $W_3^8$};
\node at (1,.8) {\tiny $M^6$};
\node at (2.8,1.5) {\tiny $W_2^8\cap N^7$};
\node at (-.8,1.5) {\tiny $W_1^8\cap N^7$};
\node at (1,-.4) {\tiny $W_3^8\cap N^7$};
\filldraw [black] (1,1) circle (1.5pt);
\end{tikzpicture}
\end{center}
The bisecting segments correspond to the double intersections of the 
corresponding $W_i^8$'s.
The nonsingular intersection form on $H^3(M^6; \R)$
comes equipped with three Lagrangian subspaces
\bea
L_1 &=& {\rm Image}\left( 
H^3(W_2^8 \cap W_3^8; \R) \to H^3(M^6; \R)
\right)\;,
\nonumber\\
L_2 &=& {\rm Image}\left( 
H^3(W_1^8 \cap W_3^8; \R) \to H^3(M^6; \R)
\right)\;,
\nonumber\\
L_3 &=& {\rm Image}\left( 
H^3(W_1^8 \cap W_2^8; \R) \to H^3(M^6; \R)
\right)\;.
\eea
The signature defect is given by Wall's non-additivity invariant 
$\mu (L_1, L_2, L_3) \in \Z$, which coincides with the Maslov index, 
\(
{\rm sign} (W^8)= {\rm sign} (W_1^8) + {\rm sign} (W_2^8) + {\rm sign} (W_3^8) 
+ \mu (L_1, L_2, L_3)\;.
\)
The interpretation is that
 we are looking at those classes of gerbes on the worldvolume which
arise from corresponding classes of gerbes in seven dimensions. 
Following the discussion similar to the case of the M2-brane above, 
but now for degree three  cohomology rather than for degree one 
homology, we have the following interpretation. 
We are looking at those gerbes whose H-fields have the 
symmetry $h_1 + h_2 + h_3=0$, $h_i \in L_i$, on the M5-brane 
worldvolume, modulo those which have the cancellation in 
pairs property, i.e. $h_1 + h_2=0=h_1 + h_3$. This then gives
a vector space as in \eqref{W}, with the $K$'s replaces by the 
$L$'s and the definition of the Maslov index is similar to 
\eqref{mu}. The contribution to the exponentiated action
 is  then of the form 
$e^{2\pi i \frac{k}{8}\mu}$.

\subsection{The M-branes and the $f$-invariant }
\label{Sec f}

We have indicated that the correct replacement for manifolds with corners 
of the eta-invariant (or $e$-invariant) 
for manifolds with boundary is the $f$-invariant of \cite{Lau}. 
In the context of M-theory this is realized in \cite{F}. Other variants
of this are also used in \cite{String}.

\medskip
We have so far been dealing with signature operators. We would like to 
consider the $f$-invariant in terms of  Dirac operators instead;
physically this is justified by  having spinors on the worldvolume,
and mathematically this is possible because of the direct relation between the 
indices of the two types of operators. In fact, as
explained in \cite{HHM} and utilized in this context in \cite{corner}, 
a Dirac operator can be viewed as 
a (generalized) signature operator. From another angle and 
explicitly, for the case of the M2-brane, we have in degree four the $L$-genus
$L_4=\tfrac{1}{3}p_1$ and the Dirac genus  $\widehat{A}_4=-\tfrac{1}{24}p_1$,
which can be directly related as $L_4=-8\widehat{A}_4$.
For the M5-brane we have degree eight components $L_8=\tfrac{1}{3^2\cdot 5}(7p_2- p_1^2)$
and $\widehat{A}_8=\tfrac{1}{2^7\cdot 3^2 \cdot 5}(7p_1^2- 4p_2)$
related as $\tfrac{1}{8}\left( \lambda^2- L_8\right)=28\widehat{A}_8$.
Therefore, expression \eqref{PF} can be written using solely the Dirac index as
\(
\exp\left[ 2\pi i k \cdot 28 \int_{W^8} \widehat{A}_8\right]\;.
\)
Note that an analogous analysis in 
degree twelve  is useful in the case 
of M-theory \cite{sig}.


\paragraph{The real and complex $e$-invariant.}
As the expression of the $f$-invariant will involve the $e$-invariant, we 
start by describing the latter in our setting, following \cite{BN} \cite{Bod1} \cite{Bod}. 
Consider the canonical Spin${}^c$ Dirac operator $D$ on our $(U, {\rm fr})$-manifold
$W^4$ with Levi-Civita connections $\nabla^{TW, LC}$ on its tangent bundle $TW^4$,
and
with boundary the M2-brane worldvolume $M^3$. 
Consider also a bundle $E$ over $W^4$ with a connection $\nabla^E$ and 
a restriction to $M^3$. Then  the reduced eta-invariant 
$\xi(D_M):= \tfrac{1}{2}(\eta + \dim \ker D_M)$ for $D_M=D|_M$
is given by 
$
\xi (D_M) \equiv \int_W Td (\nabla^{TW, LC}) \mod \Z$, 
where $Td$ is the Todd genus. The complex $e$-invariant is 
$
e_\C (M^3) \equiv \xi (D_M) + \int_W \left\{ Td (\nabla^E) - Td (\nabla^{TW, LC}) \right\}$
mod $\Z$.
By Stokes' theorem, this can be reduced to an integral over $M^3$. With the
identification of the C-field with the Chern-Simons form
\footnote{Note that an explicit proposal for viewing the C-field as an index gerbe
is given in \cite{S-gerbe}.}
 this becomes
$
e_\C (M^3) \equiv \xi (D_M) + \int_{M^3} C_3 \mod \Z$.
On the other hand, the real $e$-invariant is 
$
e_\R (M^3) \equiv \tfrac{1}{2} \langle \widehat{A}(TW^4), [W^4, M^3] \rangle 
\equiv
\tfrac{1}{2}\int_{M^3} C_3 + \xi (D_M)  \mod \Z$.
The real and complex $e$-invariants are related as follows. 
The Spin cobordism $M{\rm Spin}_3=0$ and the 
almost complex cobordism MSU${}_3=0$ are both trivial, 
and since the first Chern class of an SU-manifold is trivial, 
the Todd genus coincides with the Dirac genus, so that 
$
\tfrac{1}{2}e_\R \equiv e_\C$ mod $\Z$.
For example, for the 3-sphere, 
$e_\R (S^3)= - \frac{1}{12}$ and $e_\C (S^3)=-\frac{1}{24}$ or $\frac{11}{24}$.

\paragraph{The M2-brane with a corner and the $f$-invariant.}
We will study a representative example. 
Consider the open  M2-brane worldvolume as the manifold with boundary $S^1 \times \D^2$.
This can be lifted to four dimensions by considering it as the boundary of 
$W^4=\D^2 \times \D^2$, the product of two closed 2-disks. 
On the other hand, one can take the boundary of the open M2-brane leading to 
a corner $S^1 \times S^1$. 
Now the generator $\eta \in \pi_1^{\rm st} \cong \Z_2$ can be represented by 
the circle with its non-bounding framing.
 Using a 
Fourier decomposition,  the 
Spin${}^c$ Dirac operator has symmetric spectrum and a single zero mode, so 
$e_\C (\eta)=\frac{1}{2}$. Then, using \cite{Bod1} \cite{Bod},  the $f$-invariant 
of the corner $S^1 \times S^1$ is 
\(
f( \eta^2) \equiv \tfrac{1}{2}\frac{E_1 -1}{2} \equiv \tfrac{1}{2} \sum_n 
\sum_{d \mid n} (\tfrac{d}{3}) q^n= \tfrac{1}{2}q + \tfrac{1}{2}q^3 + \O(q^4)\;,
\label{fet}
\)

\vspace{-3mm}
\noindent where $E_1=1+ 6 \sum_n \sum_{d|n} \left( \tfrac{d}{3}\right)q^n$ 
is the modular form of weight one for
 the congruence subgroup $\Gamma_1(3)$.
Note that if $q=0$ then $f=0$, so this expression is, in some sense, nonclassical. 
This is compatible with the fact that the $f$-invariant is associated with chromatic 
level two in homotopy theory.
Expression \eqref{fet} is the $q$-expansion contribution to the effective
action of the M2-brane due to the corner.

\paragraph{The self-dual string as a string theory via the M2-brane.}
We have seen above that 
the corner structure is not merely a mathematical tool. 
We will further describe a setting where it is 
 in fact a very useful part of the 
structure of the physical theory. 
One natural question is to what extent  the self-dual 
string forms an actual string theory, in the sense of 
perturbative expansion and cobordism. The usual way that this
string is constructed does not immediately allow for such a description. 
We propose that the concept of corners naturally gives such a desirable type of string theory. 
The M2-brane has a boundary $\partial {\rm M2}=\Sigma_2$ that lies on the M5-brane. 
In order for this $\Sigma_2$ to qualify for a string theory in the sense of 
cobordism, we must have a nontrivial boundary $\partial \Sigma_2 \neq \emptyset$. 
Therefore, we take the M2-brane itself to be a manifold with corners
of codimension-2. 
This point of view, although straightforward and obvious in light
of the above discussion, does in fact solve a problem. 
This allows for the construction of (nontrivial) elliptic 
objects needed for the elliptic cohomology description of the M2-branes
\cite{tcu}. 
We will provide the corresponding details and further consequences 
elsewhere.

\paragraph{The M5-brane as a corner and the $f$-invariant.}
We now consider
 the M5-brane and concentrate on the decomposable case,
$M^6=M^3_1\times M^2_2$, where 
 $M^3_1$ and $M^3_2$ are 3-dimensional framed manifolds.
Let
$m(M^3_i)$ be any modular form of weight 2 with respect to the fixed congruence 
subgroup $\Gamma= \Gamma_1 (N)$ with
$\overline{m} (M^3_i)= m(M^3_i) - e_\C (M^3_i) \in \Z^\Gamma [[q]]$.
Then the (geometric) $f$-invariant of the product is \cite{Bod}
\(
\check{f}( M^3_1 \times M^3_2)\equiv \overline{m} (M^3_1) e_\C (M^3_2) 
\equiv - \overline{m} (M^3_2) e_\C (M^3_1)\;.
\)
In particular, the geometric $f$-invariant of a product is antisymmetric 
under exchange of the factors, in contrast to the case of the Hopf invariant. 
We will concentrate on a representative situation, which is the 
product of two 3-spheres, $M^6=S^3\times S^3$; this is the generator of the framed 
cobordism group in six dimensions. 
The study of the elliptic genus of the products of two closed disks 
$Ell(\bb{D}^4 \times \bb{D}^4)$ in M-theory and the relation elliptic cohomology 
is taken in \cite{String}.
Now consider $S^3$ as the sphere bundle $S(\cL)$ of the Hopf line bundle over $S^2$.
The framing of the base and the fiber leads to a  framing of the total space. 
The complex Adams $e$-invariant is $e_\C (S(\cL))=-\frac{1}{12}$.
Since ${\rm MSpin}_3=0$,  the framed 3-manifold $M^3$ is 
the boundary of a Spin 4-manifold $W^4$. 
This disk bundle $\mathbb{D}(S(\cL))$ represents the element 
$\nu$ in the 
homotopy groups of spheres. Then, using \cite{Bod1} \cite{Bod}, the $f$-invariant 
of the product is given by
\(
f(\nu^2) \equiv \tfrac{1}{2} \left(\frac{E_1^2-1}{12} \right)^2= \tfrac{1}{2}q^2 + 3q^3 + \tfrac{11}{2}q^4 + \O(q^5)\;.
\label{fnu2}
\)
This essentially corresponds to two copies of the M2-brane with no boundary theory,
and can be viewed as an $f$-invariant description of the cup product 
Chern-Simons theory described in \cite{cup} \cite{up}. 
Note that $f=0$ when $q=0$, so that this expression is purely modular or, in some sense, 
quantum. Expression \eqref{fnu2} is the contribution to the effective action of the M5-brane 
due to the corner in the representative case.


\subsection{M5-brane, $G_2$ holonomy and the $\nu$-invariant}
\label{Sec G2}

We consider the M5-brane worldvolume $M^6$ as 
the base of a 7-dimensional cone, which  admits a $G_2$-holonomy
structure. Topologically, a cone is a product $(M^6 \times I)/(M^6\times \{0\})$
with a metric making the size of the interval go to zero on 
one end. That is, we have a cylinder with one end collapsed to a point. 
We then take this 7-manifold to be the boundary of 
an 8-dimensional manifold, thus making $M^6$ a codimension-2 
corner and 
at the same time admitting extra geometric structures. 
A standard example is our favorite, namely $M^6=S^3 \times S^3$; 
see e.g. \cite{Cu} for an application in M-theory on $G_2$-manifolds.

\medskip
Another setting in which special holonomy arises is when we
take the 7-dimensional $G_2$-holonomy manifold to be the total space of a circle 
bundle with 
base the M5-brane worldvolume $M^6$, with ${\rm SU}(3)$-structure. 
The relation between the 3-form $\omega_3$ defining the $G_2$-structure
 in seven dimensions and the 2-form in six dimensions is explained 
generally in \cite{Hi}.
We will identify the C-field with that 3-form, i.e. $C_3=\omega_3$.

\medskip
We now make connection to the $\nu$-invariant 
\footnote{Of course this is not the generator $\nu$ encountered previously in Section \ref{Sec f}.}
of \cite{CN}. 
We consider $W^8$ to be an eight-dimensional manifold with  Spin(7)
holonomy.  For any ${\rm Spin}(8)$ bundle on $W^8$ the Euler class
for the positive and negative chirality is given by \cite{IPW}
\(
e_\pm (W^8)= \tfrac{1}{16}(p_1^2 - 4p_2^2 \pm 8e)
\label{epm}
\)
from which, using the Spin(7) structure, 
one has the following relation between degree eight genera 
$
48 \widehat{A}(W^8) + \chi (W^8) - 3 {\rm sign}(W^8)=0$.
When there is a boundary $N^7=\partial W^8$,  the
Spin(7) structure induces a boundary $G_2$-structure. In this case, 
one can consider the $\widehat{A}$-defect
$
\nu (\omega_3):=\chi (W^8) - 3 {\rm sign} (W^8) \mod 48 \in \Z_{48}$,
 which depends only on the $G_2$-structure on $N^7$.
For example, we can take $W^8=\mathbb{D}^8$ with its flat 
${\rm Spin}(7)$-structure, having a boundary 
$S^7$ with a standard $G_2$-structure $\omega_3$. Then the 
invariant is given by $\nu (\omega_3)\cong 
\chi (\mathbb{D}^8)- 3 {\rm sign}(\mathbb{D}^8) \cong 1$ mod 48. 
We propose to view this {\it canonical $G_2$-structure}, 
as minimal (nontrivial) among the set of 
$G_2$ structures; this set is isomorphic to $H^7(N^7; \pi_7(S^7))\cong \Z$.
That is, the canonical $G_2$-structure minimizes the classical topological action
for the M5-brane. This is analogous to the canonical (2-)framings, discussed earlier,
as well as canonical String structures, discussed in \cite{tcu} \cite{tcu2}.

\medskip
We now see that the $\nu$-invariant can be deduced from the 
one-loop polynomial $I_8$, and hence from (part of) the dynamics of the 
M-branes. Observe that the combination of the first two terms 
in \eqref{epm} is a multiple of the one-loop polynomial. 
This is used in \cite{KSpin} to write the one-loop 
polynomial in terms of the second Spin characteristic class $Q_2$,
as indicated in \eqref{Q2}. This can also be written as a linear combination of the
$L$-genus and the $\widehat{A}$-genus as $I_8=-\tfrac{1}{8}L -2 \widehat{A}$. 
In terms of ${\cal I}_8=\int_{W^8}I_8$, 
expression \eqref{epm} for closed $W^8$ 
takes the form 
\(
24\hspace{.5mm}{\cal{I}}_8 - \chi (W^8)=0\;. 
\)
This then allows us to have the following definition in the case when $w_4(W^8)=0$,
i.e.
when we have a Membrane structure, in the sense of \cite{top1}. 

\begin{definition} 
The ``$I_8$-defect" in the case when 
$W^8$ has a boundary with a $G_2$-holonomy structure is
\(
\frak{i}:= \chi (W^8) \mod 24 ~ \in \Z_{24}\;.
\label{iota}
\)
\end{definition}
We view this also as an 8-dimensional  analog of the canonical 
String structure in three dimensions. 
Note that the presence of a 7-dimensional boundary, allows for
a Chern-Simons interpretation of the one-loop term \cite{SSS2} 
\cite{tcu} leading to a description via higher bundles 
 \cite{FSS1} \cite{FSS2} \cite{cup}. 
We highlight that the $I_8$-defect 
\eqref{iota} depends both on 

\vspace{1mm}
\noindent {\bf 1.} the $G_2$ structure on the boundary $N^7$, and 

\vspace{1mm}
\noindent {\bf 2.} a choice of one half the first Spin characteristic class $\tfrac{1}{2} Q_1(N^7)$.

\vspace{1mm}
\noindent More precisely, this means that we provide an M-brane interpretation of 
a specific linear combination of the $\nu$-invariant
of the $G_2$-structure and the Gauss sum of quadratic refinement associated to a choice of 
$\tfrac{1}{2}Q_1(N^7)$. A somewhat analogous situation arises in the phase of the partition function 
 in \cite{NS5} (without the $\nu$-invariant interpretation).

\medskip
An explanatory remark is in order.
\footnote{We thank Johannes Nordstr\"om for illuminating discussions on this point.}
 A closed 8-manifold obtained by gluing two 8-manifolds 
with $w_4=0$ along the boundary need not necessarily have $w_4=0$. 
However, any 7-dimensional 
Spin manifold will have a Spin coboundary with $w_4=0$. 
The analogy in the simpler case of dimension four is that gluing two Spin 
4-manifolds along their boundary gives a Spin 4-manifold only if the Spin structure on the 
boundary is preserved.
If $W_1^8$ and $W_2^8$ are two different Spin coboundaries of $M^7$ with
$w_4(W^8_i) = 0$, then being able to choose $p_1(W^8_i)/4$ with equal image in
$H^4(M^7)$ is enough to show that the gluing $X^8 := W^8_1 \bigcup_M W^8_2$
has signature divisible by 8. That this is indeed the case can be shown using 
Milgram's theorem and appropriate quadratic refinements. 


\medskip
We have indicated how the Hopf invariant can be differentially refined. It is
obvious that the other invariants, including the Kervaire, $f$-invariant, 
and the $\nu$-invariant can be refined. The Kervaire invariant involves the 
differential refinement of the Wu classes, the $f$-invariant involves
essentially refined Chern-Simons theory, and the $\nu$-invariant 
involves refining the 3-form to a gerbe with connection, again 
obvious from Chern-Simons theory, along the lines of the above references.

\vspace{5mm}
\noindent
{\bf \large Acknowledgements}

\vspace{1mm}

\noindent This research  is supported by NSF Grant PHY-1102218.
The author thanks AIM, Palo Alto, for hospitality and Matthew Ando and 
Mike Hill for useful discussions.   He also thanks 
Ralph Cohen and Johannes Nordstr\"om for helpful comments.
The author thanks IHES, Bures-sur-Yvette, for hospitality in Summer 2013.


\end{document}